\documentclass[aps,floats,showpacs,twocolumn]{revtex4}

\usepackage[dvips]{color}     
\usepackage{epsfig,graphicx}

\newcommand{\be}{\begin{equation}}
\newcommand{\ee}{\end{equation}}
\newcommand{\bea}{\begin{eqnarray}}
\newcommand{\eea}{\end{eqnarray}}
\newcommand{\beas}{\begin{eqnarray*}}
\newcommand{\eeas}{\end{eqnarray*}}
\newcommand{\bt}{\begin{tabular}}
\newcommand{\et}{\end{tabular}}
\newcommand{\ba}{\begin{array}}
\newcommand{\ea}{\end{array}}

\newcommand{\noi}{\noindent}
\newcommand{\drm}{{\rm d}}
\newcommand{\erm}{{\rm e}}

\begin{document}


\title{{\em Pipe effect} in viscous liquids}

\author{V. Capano}
\affiliation{Dipartimento di Scienze Fisiche, Universit\`{a} di
Napoli ``Federico II'', Complesso Universitario di Monte S.
Angelo, via Cinthia, I-80126 Naples, Italy}
\author{S. Esposito}
\email{Salvatore.Esposito@na.infn.it}%
\affiliation{Dipartimento di Scienze Fisiche, Universit\`{a} di
Napoli ``Federico II'' \\ and Istituto Nazionale di Fisica
Nucleare, Sezione di Napoli, Complesso Universitario di Monte S.
Angelo, via Cinthia, I-80126 Naples, Italy}
\author{G. Salesi}
\email{Giovanni.Salesi@unibg.it} %
\affiliation{Universit\`a Statale di Bergamo, Facolt\`a di
Ingegneria, viale Marconi 5, I-24044 Dalmine (BG), Italy \\ and
Istituto Nazionale di Fisica Nucleare, Sezione di Milano, via
Celoria 16, I-20133 Milan, Italy}

\begin{abstract}
\noindent A detailed experimental and theoretical study has been
performed about a phenomenon, not previously reported in the
literature, occurring in highly viscous liquids: the formation of
a definite pipe structure induced by the passage of a heavy body,
this structure lasting for quite a long time. A very rich
phenomenology (including mechanical, optical and structural
effects) associated with the formation of the pipe has been
observed in different liquids. Actually, the peculiar dynamical
evolution of that structure does not appear as a trivial
manifestation of standard relaxation or spurious effects. In
particular we have revealed different time scales during the
evolution of the pipe and a non-monotonous decreasing of the
persistence time with decreasing viscosity (with the appearance of
at least two different maxima). A microscopic model consistent
with the experimental data, where the pipe behaves as a
cylindrical dielectric shell, has been proposed. The general time
evolution of the structure has been described in terms of a simple
thermodynamical model, predicting several peculiarities
effectively observed.

\end{abstract}


\maketitle



\section{Introduction}

\noindent One of the most active area of research in soft
condensed matter physics, physical chemistry, materials science
and biophysics is the study of the dynamics of complex systems and
their relationship to the structure. Among such complex systems, a
special place is occupied by hydrogen-bonding liquids and their
mixtures \cite{HBond1} \cite{HBond2}, due to their extreme
prominence in different biological and technological processes.
This is especially true for glycerol (C$_3$H$_8$O$_3$), for which the
presence of three hydroxyl groups per molecule makes it a
particularly rich and complex system for the study of hydrogen
bonded fluids. Glycerol, indeed, has been the subject of
considerable and long-standing scientific interest \cite{Glyc12}
due to its complex nature.

For example, the peculiarities of the intermolecular interaction
of glycerol with water via hydrogen bonds form the basis of the
valuable hydration properties of glycerol, that are widely applied
in the pharmaceutical, cosmetics and food industries. Just to
quote typical examples, glycerol has been employed as a
cryoprotector \cite{cryopro}, depending upon the changes
of the parameters of the phase transitions of water in the
presence of glycerol, but it has been also the focus of study by
researchers in cryopreservation \cite{cryopre}.

On the other hand, glycerol is an excellent glass former, and has
been extensively studied experimentally \cite{glassformer} in
connection with attempts to understand the nature of the glass
transition. Understanding the liquid-glass transition and its
related dynamics, in fact, is one of the most important and
challenging problems in modern condensed matter physics, and
glycerol and its mixtures with water are widely used as models to
study the cooperative dynamics, glass transition phenomena and
scaling properties in complex liquids.

The dielectric properties of hydrogen-bonded liquids are, as well,
of key interest, because such liquids generally shows abnormal
dielectric behavior, which is not observed in a
non-hydrogen-bonded liquid. In fact, with some exception (such as
acetic acid), the static dielectric constant of a hydrogen-bonded
liquid is generally larger than that of a normal polar one, mainly
because of the regular alignment of a dipolar molecule in the
hydrogen-bonded cluster. This applies especially to glycerol,
where the existence of long- and short-ranged forces, and a large
variety of possible molecular conformations, leads to important
dynamics on a variety of time scales.

All these facts evidently motivate, on one hand, the large
research effort about hydrogen-bonded liquids and, in particular,
glycerol, but also urge to study further such systems, searching
for possible novel phenomena that put some other light on their
interesting physical properties. In the present paper we
extensively report just on an apparently new effect of such kind
observed in glycerol that, seemingly, has not been considered
before (at least in the published literature).

The starting observation is strictly related to the standard
determination of the viscosity by means of the method of the
falling sphere. Although not easily seen with the naked eye, after
the falling of the metal sphere in glycerol at common
temperatures, a pipe appears in the viscous liquid that persist
for some long time. At a first glance, such a not surprising
effect seems to be easily explained in terms of some relaxation
processes occurring in highly viscous media. However, the lacking
of an apparent microscopic structure in liquids, though highly
viscous, similar to solids seems mainly to disfavor the simple
explanation envisaged above, while requiring more observations on
the phenomenon. Indeed, such observations do reveal a very rich
phenomenology, pointing out that the effect observed is, quite
surprisingly, not at all trivial.

We have thus started a set of appropriate experiments aimed at
collecting all the relevant phenomenology (both qualitative and
quantitative), upon which a theoretical explanation of the
phenomenon, though preliminary, may be consistently built. The
results of our large study are reported here. After a detailed
description in Sections II of all the experimental observations
obtained, together with the report on the results of data fitting
and their interpretation, in Section III we give possible
theoretical interpretation of the phenomena observed. In
particular, we propose a possible microscopic interpretation of
the occurrence of the pipe effect, discussing quite in detail the
corresponding theoretical model, and discuss and solve a thorough
thermodynamical model describing the evolution of the pipe,
deducing some of the properties observed. Finally, in Section IV,
we summarize the results obtained and give our conclusions and
outlook.

\begin{table}
\begin{center}
\footnotesize
\begin{tabular}{|c|ll|lll|lll|}
\hline & Property &  & & Glycerol
 & & & Water & \\
\hline $M$ & Molecular weight (g/mol) & & & 92.09
& & & 18.01 & \\
$\rho$ & Density\footnote{At $25 ^{\mathrm o}$C, $101.325$kPa.}
(Kg/m$^3$)
& & & 1258.02 & & & 997.05 & \\
$\eta$ & Viscosity\footnote{At $20 ^{\mathrm o}$C, $101.325$kPa.}
($10^{-3}$ Pa $\!\cdot\!$ s)
& & & 1410 & & & 1.0016 & \\
$\sigma$ & Surface tension\footnote{At $25 ^{\mathrm o}$C.}
($10^{-3}$ N/m)
& & & 64.00 & & & 71.98 & \\
 \hline
$\beta$ & Compressibility, isothermal
& & &  & & & & \\
 & ($10^{-10}$/Pa)
& & & 1 $\div$ 0.1\footnote{See L.J. Root and B.J. Berne, J. Chem.
Phys. {\bf 107} (1997) 4350.}
& & & 4.599$^c$ & \\
$c$ & Specific heat$^a$ (J/Kg $\!\cdot ^{\mathrm o}$K)
& & & 2406 $\div$ 2425 & & & 4181.9 & \\
$\kappa$ & Thermal conductivity$^c$
& & &  & & &  & \\
 & (W/m $\!\cdot ^{\mathrm o}$K)
& & & 0.285 & & & 0.610 & \\
$\gamma$ & Thermal expansion coefficient$^c$
& & &  & & &  & \\
 & ($10^{-6}$/$^{\mathrm o}$K)
& & & 615 & & & 253 & \\
 \hline

$\mu$ & Electric dipole moment
& & &  & & &  & \\
 & (D = $3.33564 \!\cdot\! 10^{-30}$ C $\!\cdot\!$ m)
& & & 2.68\footnote{See K.D. Cook, P.J. Todd and D.H. Friar,
Biomed. Environ. Mass Spectrom. {\bf 18} (1989) 492.}
 & & & 2.95\footnote{At $27 ^{\mathrm o}$C} & \\
$\alpha$ & Polarizability volume
& & &  & & &  & \\
& ($10^{-30}$ m$^3$) & & & 8.13 & & & 1.470 & \\
$\varepsilon$ & Dielectric constant$^c$ & & & 40.10 & & & 78.4 & \\
$n$ & Refractive index\footnote{At $25 ^{\mathrm o}$C, $\lambda =
589.26$nm}
& & & 1.47399 & & & 1.33286 & \\
\hline
\end{tabular}
\caption{Main physical properties of liquid glycerol
\cite{glycerolprop} and water \cite{waterprop}.
\newline} \label{tableprop} \vspace{-0.7truecm}
\end{center}
\end{table}

\section{Phenomenology}

\begin{figure}
\begin{center}
\epsfxsize=8cm %
\epsffile{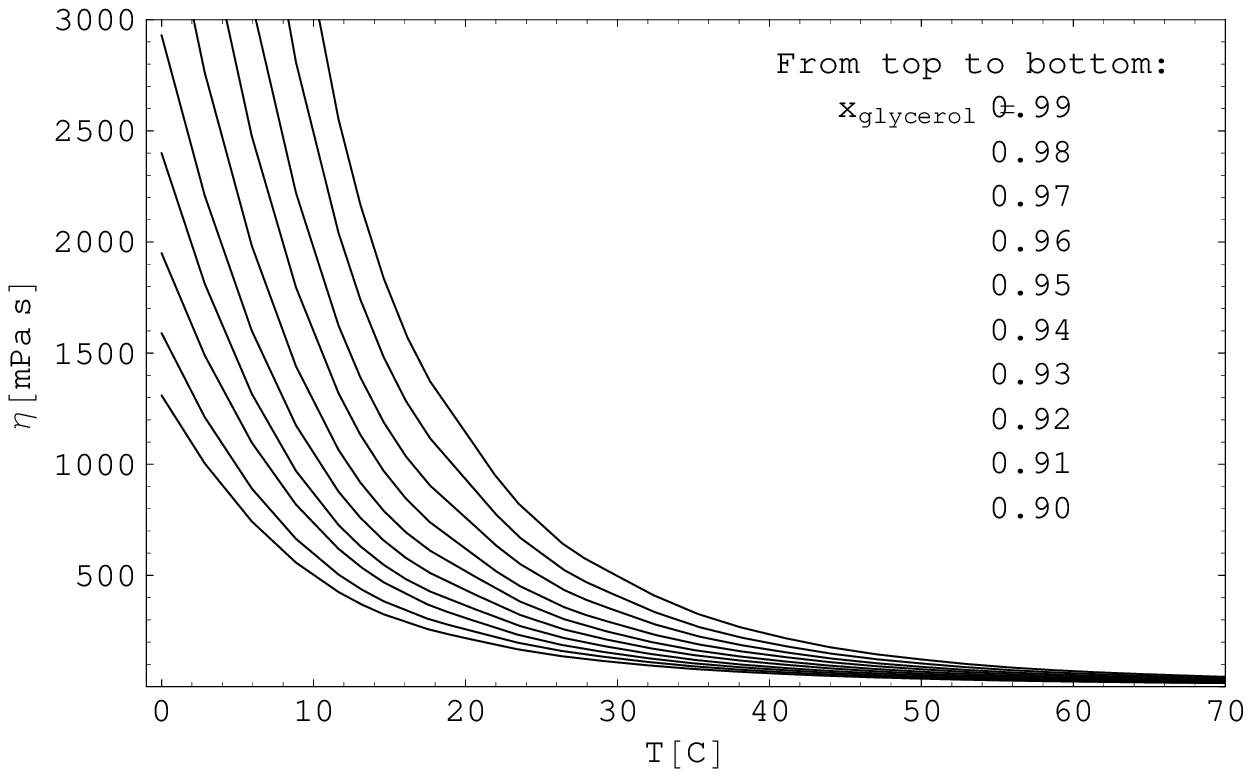} a) \\ ${}$ \\
\epsfxsize=8cm %
\epsffile{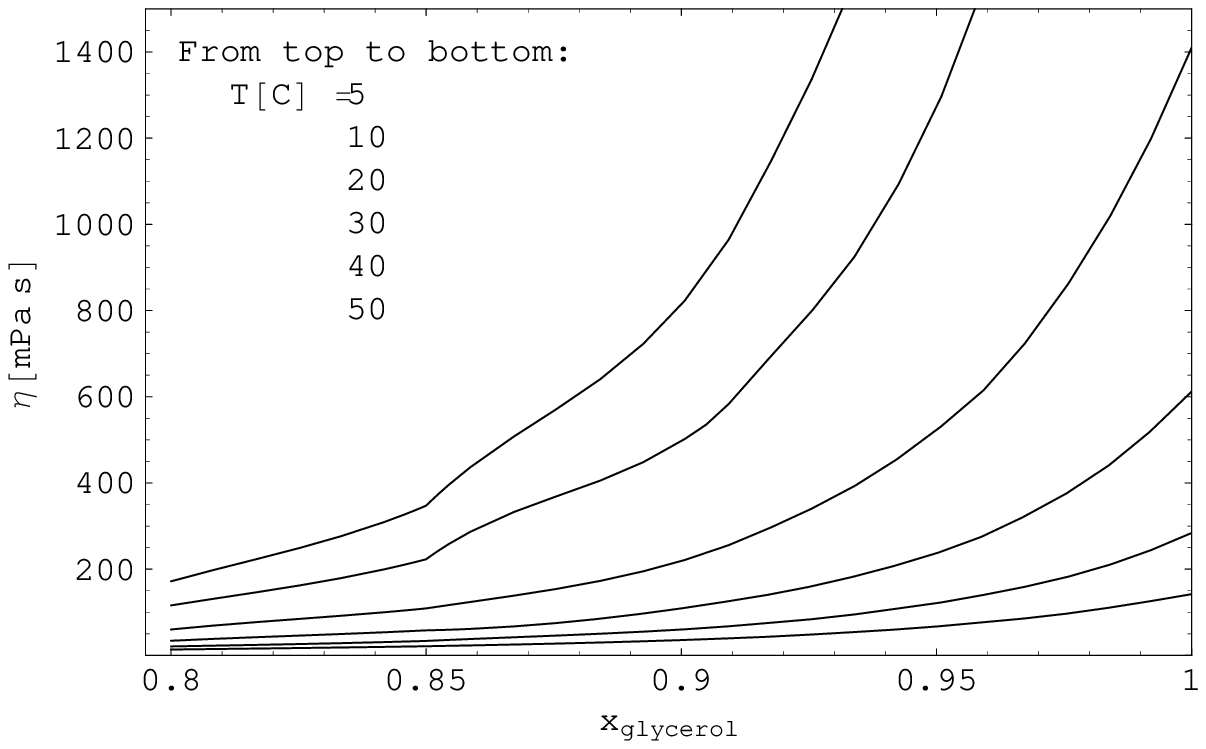} b)
\caption{Dependence of the viscosity $\eta$ of glycerol-water
solutions on (a) temperature $T$, and (b) concentration
$x_{\rm glycerol}$ of glycerol. \hfill\break [The curves have been
elaborated from numerical data given by ``The Dow Chemical
Company'' on the website
{\small http://www.dow.com/glycerine/resources/physicalprop.htm}.]
\newline} \label{etatx}
\end{center}
\vspace*{-1.7truecm}
\end{figure}

\begin{figure}
\begin{center}
\epsfxsize=8cm %
\epsffile{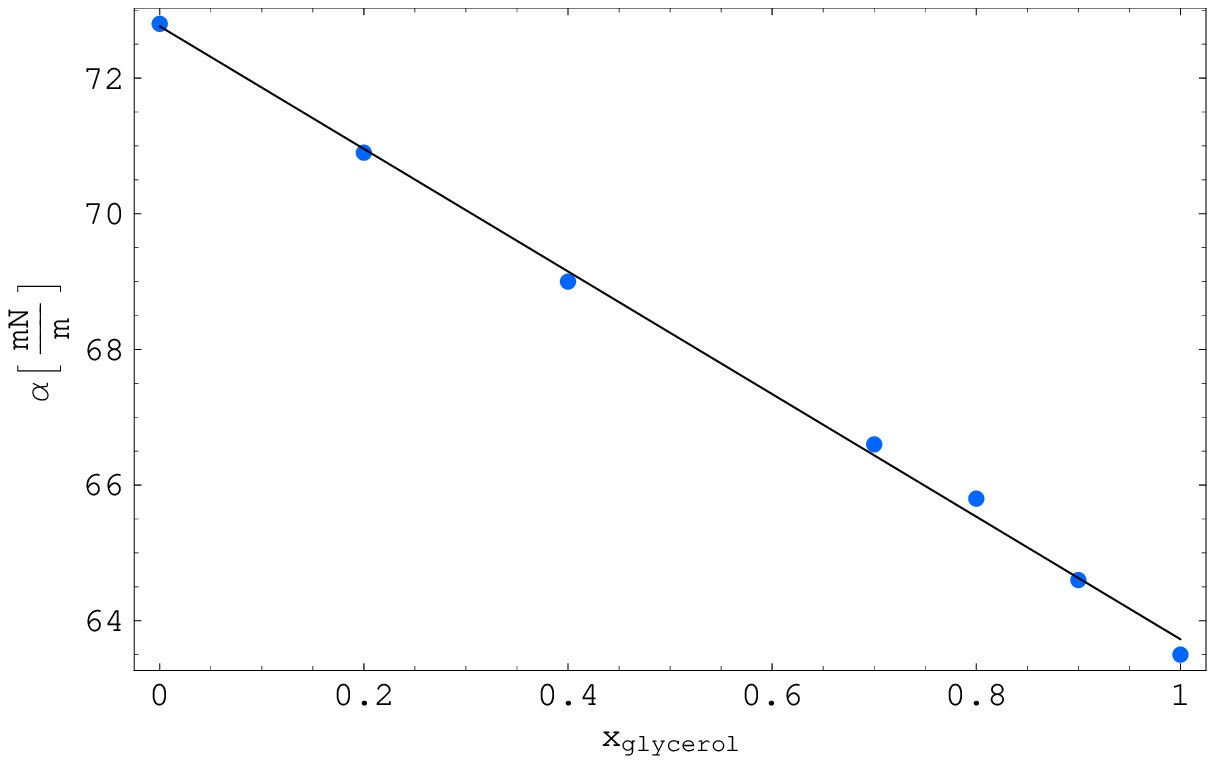} a) \\ ${}$ \\
\epsfxsize=8cm %
\epsffile{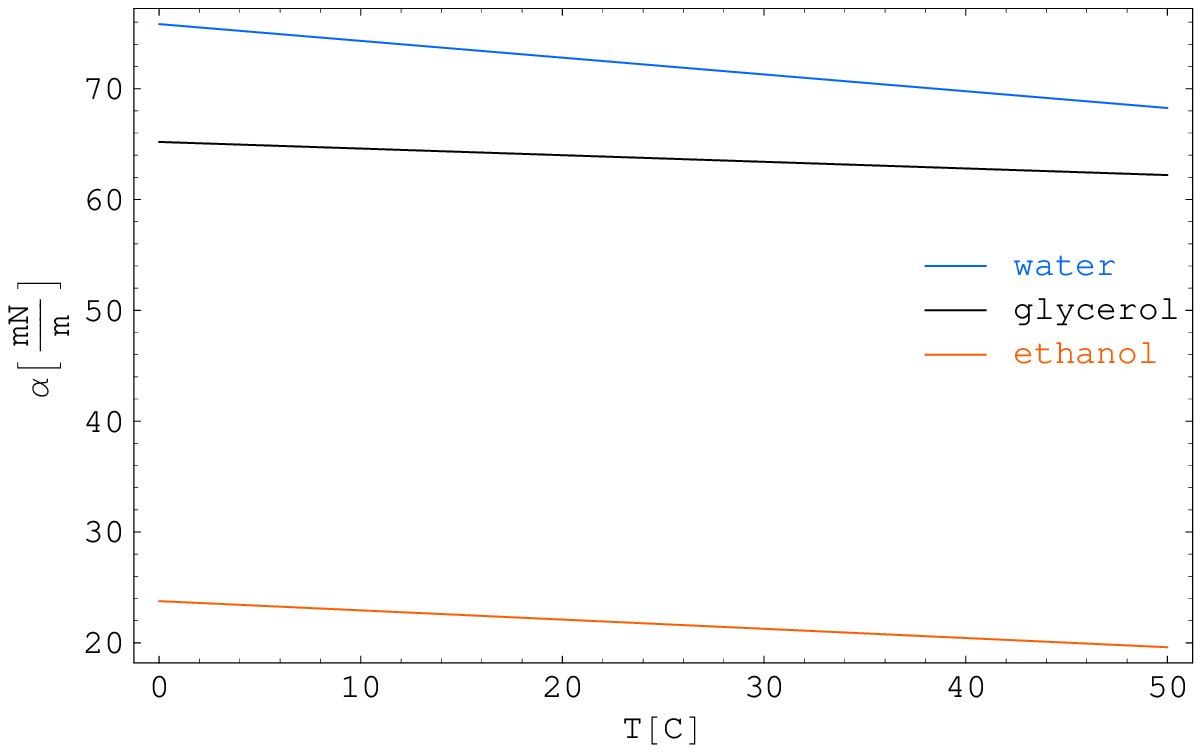} b)
\caption{a) Dependence of the surface tension $\sigma$ of
glycerol-water solutions on the concentration $x_{\rm glycerol}$ of
glycerol ($T=20 ^{\rm o}$C). The data are approximated by the
relation $\sigma = 72.76 - 9.04 x_{\rm glycerol}$.
\newline [Elaboration on the numerical data given by N.R. Morrow,
``Fundamentals of reservoir surface energy as related to surface
properties, wettability, capillary action, and oil recovery from
fractured reservoirs by spontaneous imbibition'', Quarterly Report
DE-FC26-03NT15408, from the U.S. DOE's Office of Scientific and
Technical Information website {\small http://www.osti.gov}.]
\newline b) Dependence of the surface tension $\sigma$ of pure
liquids (water, glycerol and ethanol) on the temperature $T$. The
curves correspond to the following fitting relations:
$\sigma_{\rm water} = 72.8 - 0.1514 (T-20)$, $\sigma_{\rm glycerol} = 64.0
- 0.0598 (T-20)$, $\sigma_{\rm ethanol} = 22.1 - 0.0832 (T-20)$.
\newline [Elaboration on the numerical data reported on the website
{\small http://www.surface-tension.de} della  DataPhysics
Instruments.] \newline } \label{sigma}
\end{center}
\vspace{-0.7truecm}
\end{figure}

\noindent In the experiments performed, we have employed different
commercial viscous liquids, glycerol (C$_3$H$_8$O$_3 \geq 99.5
\%$, water content $\leq 0.1 \%$), ethanol (C$_2$H$_6$O $\geq 96
\%$) and castor oil (essentially pure), as delivered, while, for
the mixtures with water, a doubly distilled water was used. We
have focused our attention particularly to glycerol, whose main
physical properties (and, for comparison, those of water) are
reported in Table \ref{tableprop} for a reference temperature. For
practical and interpretational uses, it is also helpful to take
into account the dependence of the most directly relevant
properties of the viscous liquids used on temperature $T$ (always
given in Celsius degrees $^{\mathrm o}$C) and concentration $x$;
the corresponding graphs are showed in Figures \ref{etatx},
\ref{sigma}. For the sake of simplicity, in the following we will
refer to ``pure'' glycerol when no solute (water or ethanol) has
been added; we have checked that, in such case, the concentration
of glycerol is about $x=0.99$ (mainly due to pre-existing
impurities), that will then be our reference value.

The several experiments carried out consisted, basically, always
in inducing the formation of the pipe in the viscous liquid, just
by means of the falling (from the rest) of steel spheres of
different diameters, and then observing various properties about
the formation and evolution of the pipe, or measuring different
properties of the pipe itself. For obvious reasons, we have taken
some care in keeping constant and uniform the relevant parameters
(pressure, temperature, concentration, etc.) of the probing
liquids, contained in given graduated tubes (heights ranging from
15 to 30 cm, diameters from 2 to 5 cm).

While the formation of the pipe follows closely the falling of the
sphere (it, is, practically, instantaneous), its disappearance
after some time is not, in the given experimental conditions,
unambiguous. In order to keep the errors on the persistence time
of the phenomenon under control, we have always adopted the same
protocol to mark the ``complete'' disappearance of the pipe (that
is, when the luminosity of the pipe with respect to the bulk
liquid falls below a given value, necessarily different from
zero).

For the study of the optical properties of the pipe, we have used
standard filament and halogen lamps, and a common He-Ne laser.

The properties of the pipe have been studied in four different
kinds of viscous liquids, namely: pure glycerol, glycerol/water
mixtures and glycerol/ethanol mixtures with different
concentrations, and pure castor oil. The motivation for this
choices lies in their structural properties and, in particular, on
the values of two main quantities of those substances, that are
viscosity and surface tension. Indeed, glycerol has very high
viscosity ($\eta_{\rm glycerol} = 1410 \times 10^{-3}$ Pa $\!\cdot\!$ s at
$20^{\mathrm o}$C) and surface tension
($\sigma_{\rm glycerol} = 63.4 \times 10^{-3}$ N/m at $20^{\mathrm
o}$C), while water and ethanol have a low viscosity ($\eta_{\rm water}
= 1.002 \times 10^{-3}$ Pa $\!\cdot\!$ s and $\eta_{\rm ethanol} = 1.200
\times 10^{-3}$ Pa $\!\cdot\!$ s at $20^{\mathrm o}$C) but high
($\sigma_{\rm water} = 72.8 \times 10^{-3}$ N/m at $20^{\mathrm o}$C)
and intermediate ($\sigma_{\rm ethanol} = 22.8 \times 10^{-3}$ N/m at
$20^{\mathrm o}$C) surface tension, respectively, so that, with
different mixtures, at least the dependence on viscosity and
surface tension of pipe formation and evolution may be
discriminated. The use, in further experiments, of castor oil too,
which have an high viscosity, comparable to that of pure glycerol
($\eta_{\rm castor oil} = 986 \times 10^{-3}$ Pa $\!\cdot\!$ s), but a
completely different molecular structure, may shed some light on
the influence of the structure on the phenomenon.

In the following we will report schematically the different
observations obtained, appropriately grouped on the basis of their
physical content.

\subsection{Qualitative observations}

\subsubsection{Pipe formation}

F1. The pipe is generated after the falling of spheres of
different diameters in pure glycerol, irrespective of the starting
falling conditions (from outside the tube or from inside the
glycerol), the diameters (from 2 mm to 12 mm) and the shape
(cylinder or prism with different basis) of the tubes.

F2. The pipe does not form (in any kind of tubes, in vertical or
inclined positions) if, before falling, the spheres are deposited
in pure glycerol for a very long time (greater than five hours).

F3. The pipe does not form if, instead of falling steel spheres, a
(macroscopic) air bubble is used during its reclimbing motion in
pure glycerol.

F4. The pipe is generated (in a very visible way) if, instead of
falling steel spheres, water droplets on the bottom of the tube
are used in their reclimbing motion in pure glycerol. The pipe,
however, deforms very rapidly (water is soluble in glycerol) and,
after a given time, all the water is absorbed by glycerol, leaving
no track of its passage.

F5. The pipe is generated even in non-pure glycerol, that is in
glycerol/water mixtures (with concentration as low as
$x_{\rm glycerol} = 0.80$) and in glycerol/ethanol
mixtures\footnote{Note that even in glycerol/ethanol mixtures a
small content of spurious water is present, not lower than $1\%$,
mainly due to to the content of water in the ``pure'' glycerol
employed.} (with concentration as low as $x_{\rm glycerol} = 0.90$).

F6. When observing the projection (see the point O4) of the pipe
generated in glycerol/water mixtures with an high concentration of
water (around $8\%$ in weight of water), just after the falling of
the sphere the pipe appears to be not homogeneous but as formed by
alternatively bright and dark vertical strips. The passage of the
reclimbing mass, considered at point S3, compact such strips and
the pipe becomes homogeneous.

F7. The pipe generated in glycerol/ethanol mixtures, for any
concentration considered, appears (see the point O1) to have a
more pronounced cylindrical structure with respect to its
formation in pure glycerol or in glycerol/water mixtures (the pipe
appears, effectively, as a ``tube immersed'' in the bulk liquid).

F8. The pipe is generated by falling spheres even in castor oil,
although in a less marked way.

\begin{figure}[t]
\begin{center}
\begin{tabular}{ccc}
\epsfxsize=3.5cm %
\epsfysize=2.5cm %
\epsffile{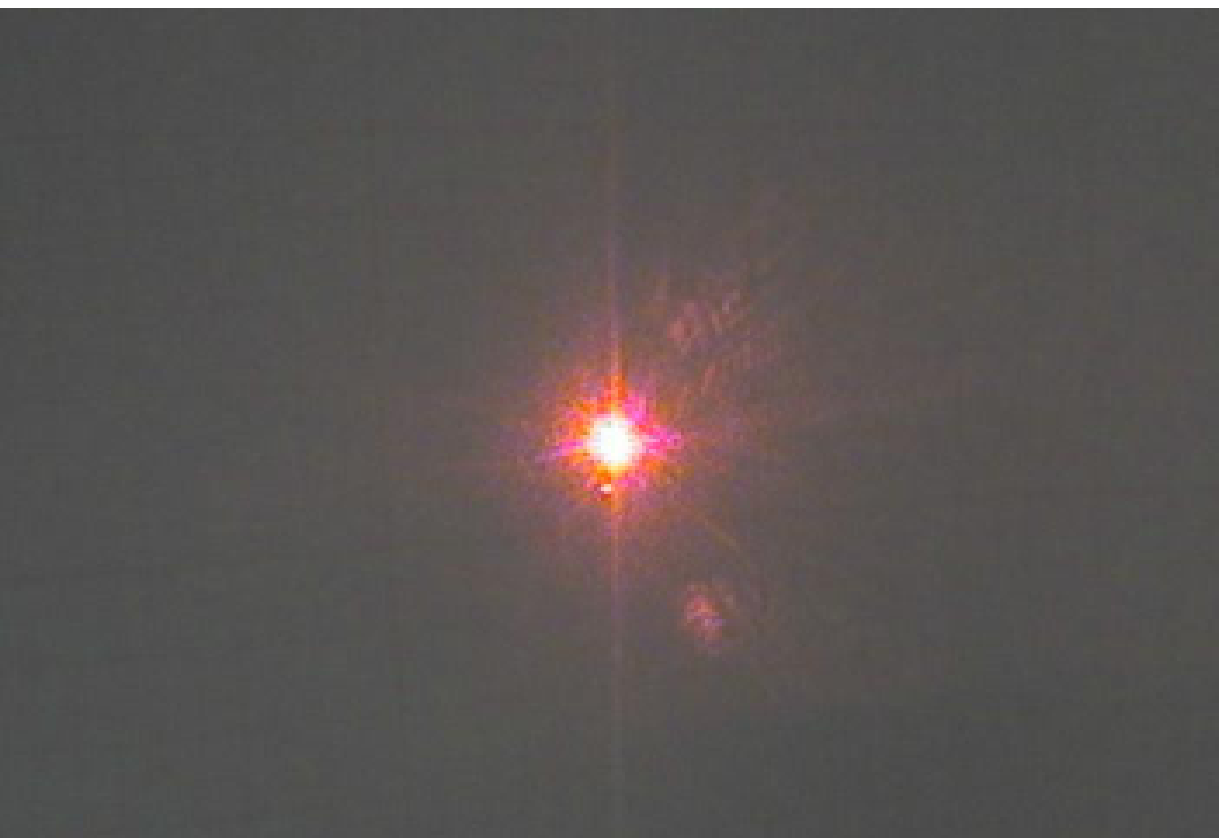} a) & &
\epsfxsize=3.5cm %
\epsfysize=2.5cm %
\epsffile{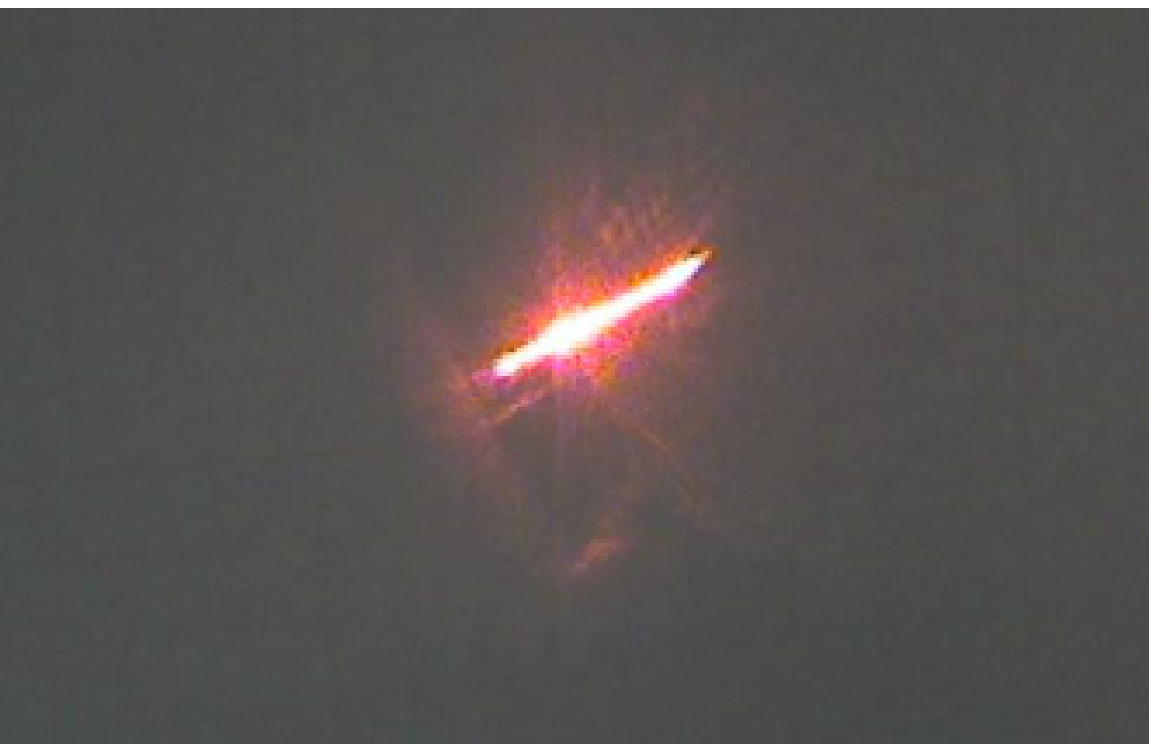} b) \\
\epsfxsize=3.5cm %
\epsfysize=2.5cm %
\epsffile{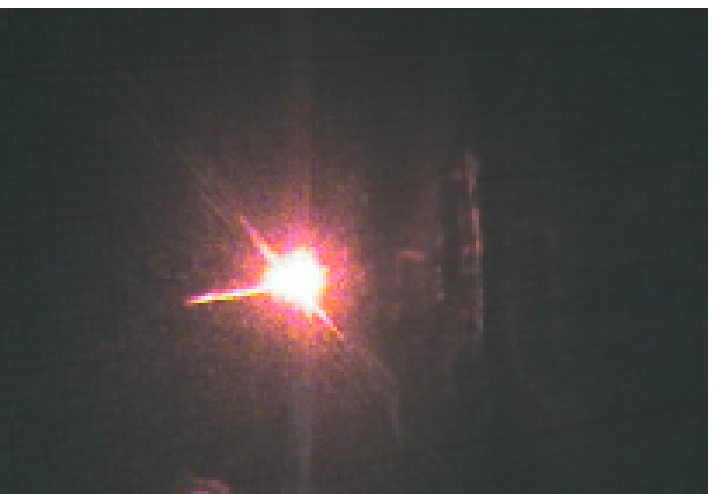} c) & &
\epsfxsize=3.5cm %
\epsfysize=2.5cm %
\epsffile{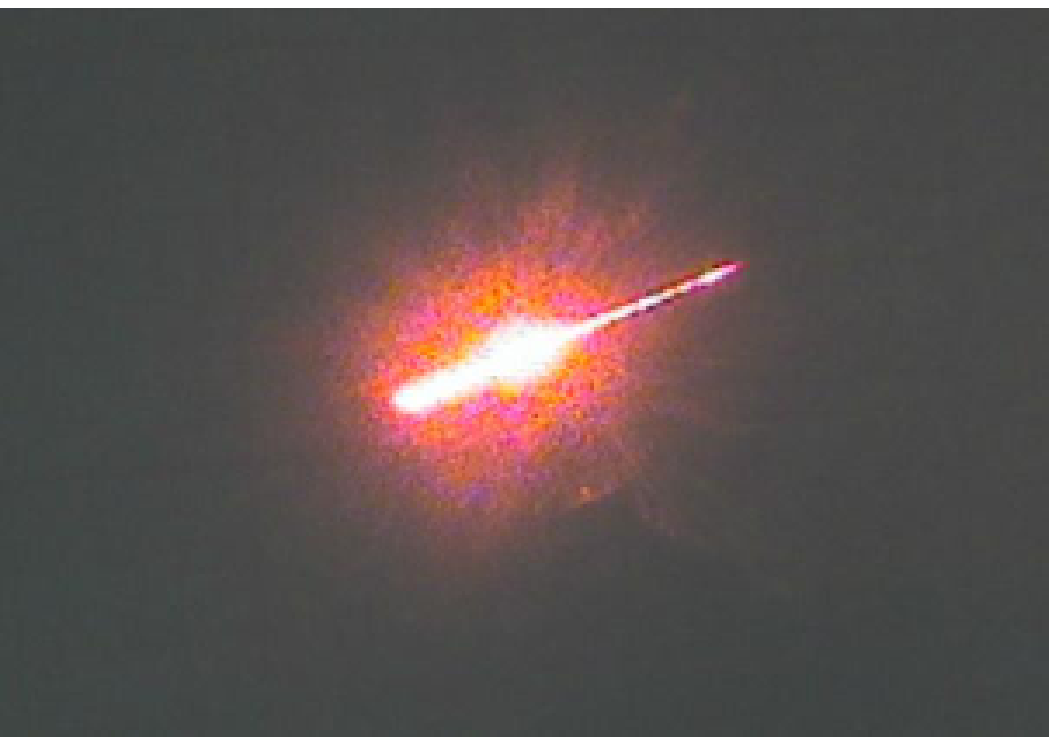} d) \\
\epsfxsize=3.5cm %
\epsfysize=2.5cm %
\epsffile{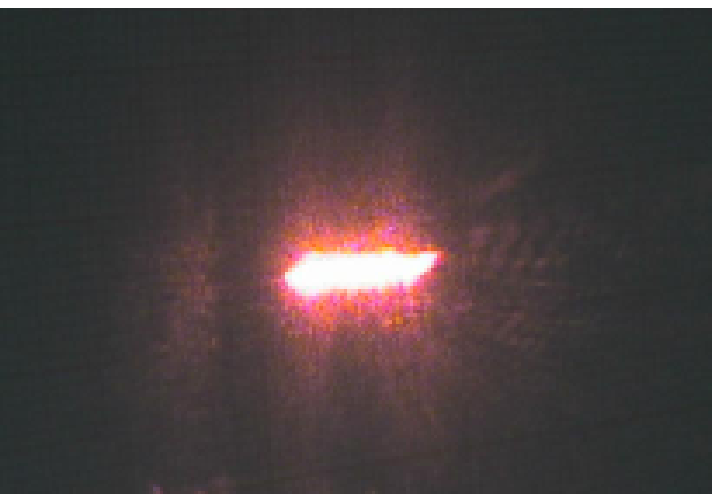} e) & &
\epsfxsize=3.5cm %
\epsfysize=2.5cm %
\epsffile{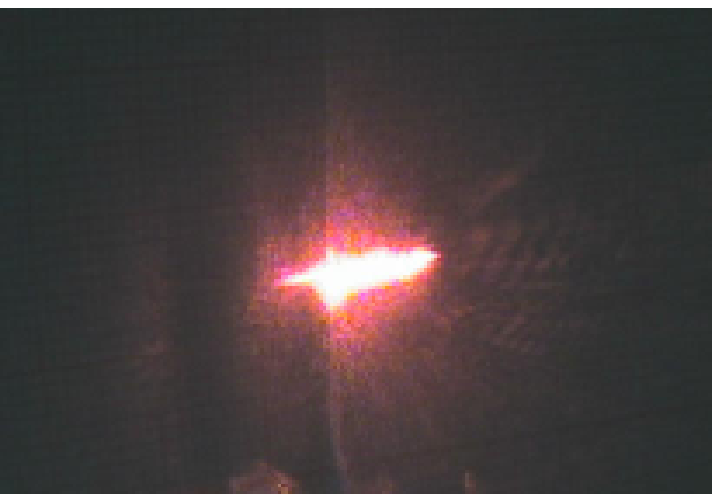} f) \\
\epsfxsize=3.5cm %
\epsfysize=2.5cm %
\epsffile{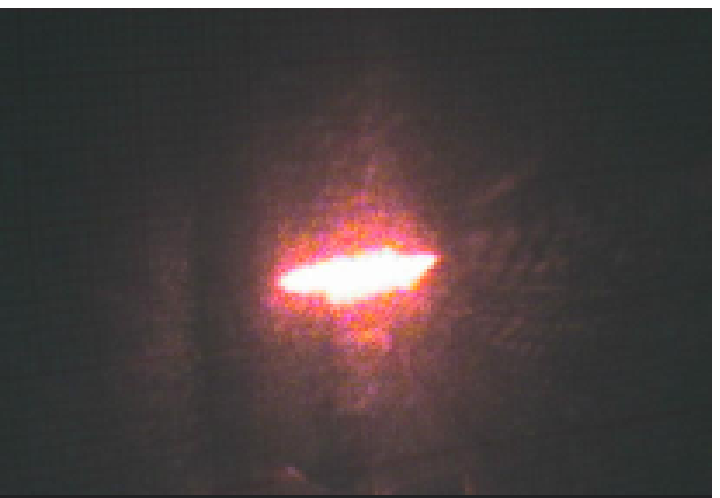} g) & &
\epsfxsize=3.5cm %
\epsfysize=2.5cm %
\epsffile{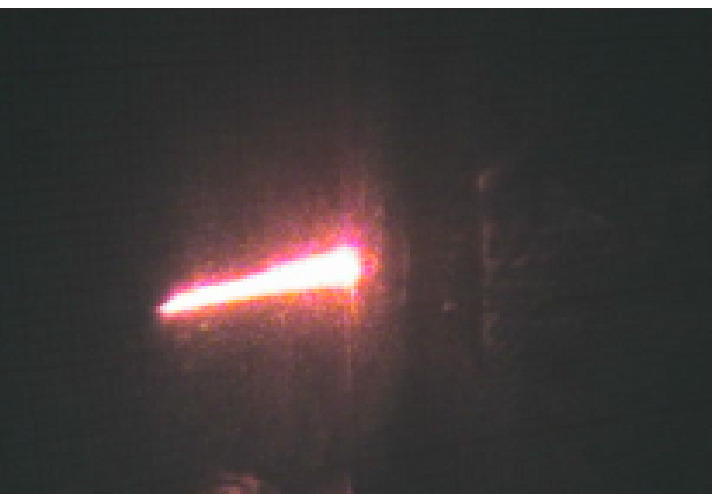} h)
\end{tabular}
\caption{Scattering of a laser beam from a pipe generated in
glycerol (O8 effect, see the text). \newline} \label{figO8}
\end{center}
\vspace{-0.7truecm}
\end{figure}

\subsubsection{Dynamics of pipe formation}

D1. Just after the falling of the sphere in pure glycerol, the
diameter of the pipe generated coincides with that of the sphere
but, very rapidly, the pipe grows thinner till its diameter takes
a stationary value (largely lower than the diameter of the
sphere). Such a thinning takes place by means of liquid reclimbing
around the pipe towards the top of the tube rather than by means
of a narrowing/absorption process. (see also the point S3.)

D2. Just after the falling of spheres with large diameters (around
10 mm), a double pipe structure is observed: a pipe with lower
diameter is inside a larger pipe.\footnote{Illuminated by normal,
not laser, light (see the point O4), the smaller pipe appears on a
screen with a darker shadow with respect to that of the larger
one, though always darker with respect to the bulk glycerol.}
Quite rapidly, however, the outer pipe gets narrower till it
coincides with the inner one.

D3. For a sphere put at rest in pure glycerol for a not extremely
large time (up to 4-5 hours), that is allowed to fall along one
side of an inclined parallelepiped (small inclinations), the
glycerol flow gliding from the front to the back of the falling
sphere is observed\footnote{With an appropriate illumination with
normal light.}, the flow closing behind the sphere. In such a
case, the pipe is not perceptible (see also the point F2).

D4. In the same conditions of point D3, but putting the sphere at
rest in glycerol for extremely long times (about one day), the
glycerol flow is barely perceptible, while for longer times it is
completely unobservable. In any case, a pipe is not generated (see
also the point F2).

\subsubsection{Size and shape of the pipe}

G1. The upper part of the pipe (in pure glycerol) is
funnel-shaped, while rapidly narrowing to a cylinder towards the
bottom.

G2. The sizes of pipes generated by differently sized spheres are
not proportional to the diameters of the spheres. For example, by
doubling the diameter of a falling sphere, the pipe generated is
only slightly larger.

G3. In glycerol/water and glycerol/ethanol mixtures, the shape of
the pipe is not regularly cylindrical (with a definite straight
line axis) as in the case of pure glycerol.

G4. For a given sized sphere, the pipe generated in castor oil has
a smaller diameter with respect to that generated in glycerol (or
glycerol mixtures).

\subsubsection{Pipe evolution}

E1. The persistence time (time from the formation of the pipe till
its disappearance) depends on the concentration of glycerol, when
glycerol/water and glycerol/ethanol mixtures are used: by lowering
the concentration of glycerol, the persistence time gets shorter
(see quantitative results below).

E2. Before the pipe disappears, it gets deformed (its shape is no
more cylindrical with a straight line axis), such a deformation
apparently depending on the (horizontal and vertical) sizes of the
tubes employed.

E3. Before the pipe disappearance, its deformation in glycerol/water
and glycerol/ethanol mixtures is analogous to that in pure
glycerol, but more accentuated (for lower glycerol concentrations).

E4. The persistence time of the pipe generated in castor oil is
much shorted with respect to (pure or mixed) glycerol (see
quantitative results below).

\subsubsection{Mechanical effects}

M1. A mechanical action may be performed upon the pipes generated
in pure glycerol without breaking them (they can be shifted,
winded one around another, and so on).

M2. During the reclimbing motion of a (macroscopic) air bubble
generated at the bottom of a pipe with diameter lower than that of
the bubble, the pipe widens out, taking again its original shape
after the passage of the bubble.

M3. The pipe generated in pure glycerol may be mechanically
deformed, generating novel pipes through bifurcations. In particular,
it is possible to ``close'' the pipe by ``transporting'' backward
(that is, in the direction opposite to that of formation)) its
surface. In such a case, however, although the pipe is
re-absorbed, a perturbed crater-like zone persists (for some time)
on the free surface of glycerol.

\begin{figure}
\begin{center}
\begin{tabular}{ccc}
\epsfxsize=3.5cm %
\epsfysize=2.5cm %
\epsffile{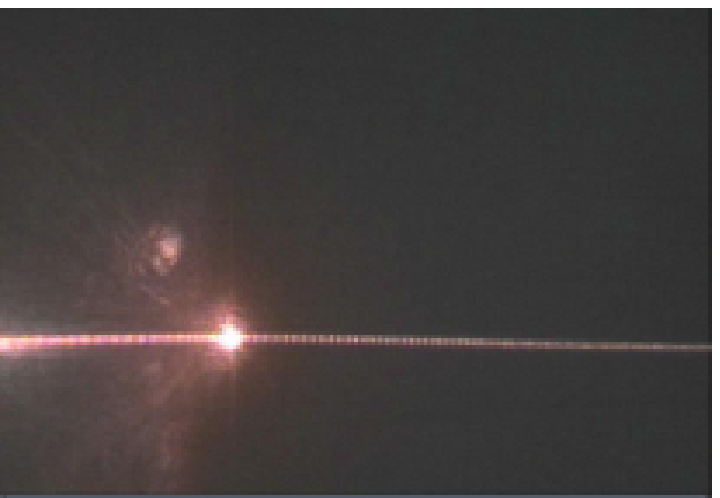} a) & &
\epsfxsize=3.5cm %
\epsfysize=2.5cm %
\epsffile{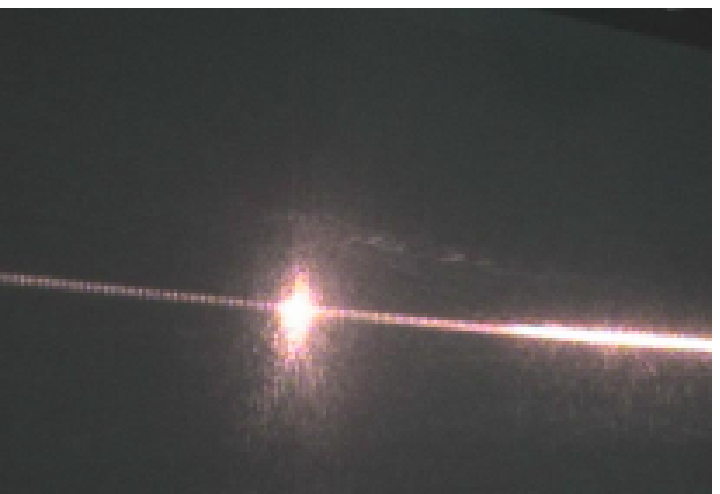} b) \\
\epsfxsize=3.5cm %
\epsfysize=2.5cm %
\epsffile{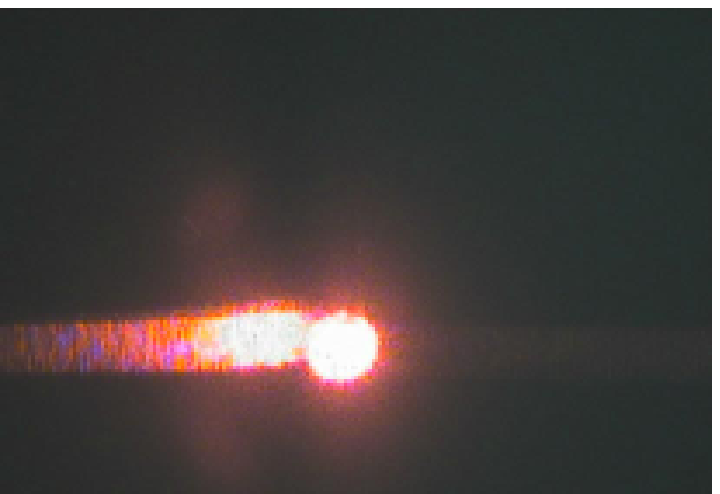} c) & &
\epsfxsize=3.5cm %
\epsfysize=2.5cm %
\epsffile{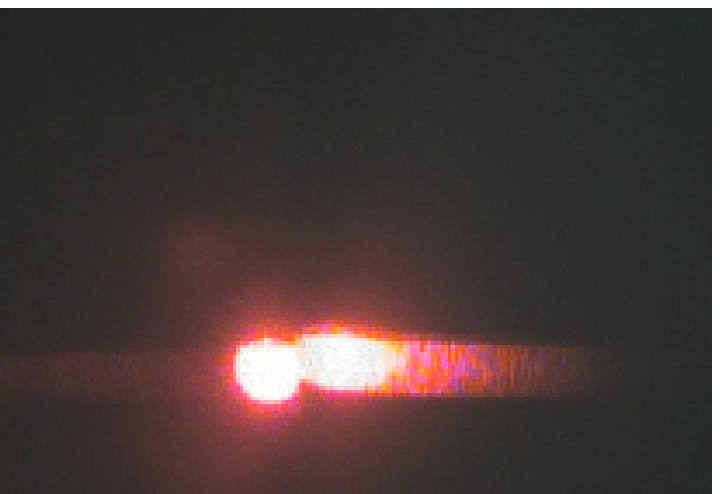} d) \\
\epsfxsize=3.5cm %
\epsfysize=2.5cm %
\epsffile{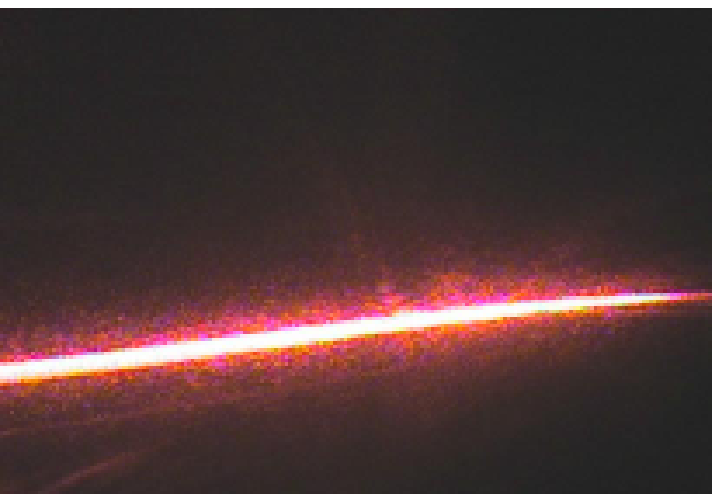} e) \\  &
\end{tabular}
\caption{Scattering of a laser beam from a glass tube immersed in
glycerol and filled with (a,b) water or (c,d) glycerol (O8 effect,
see the text). In e): particular of the scattered beam (tube
filled with water or glycerol). \newline} \label{figO8b}
\end{center}
\vspace{-0.7truecm}
\end{figure}

\subsubsection{Optical effects}

O1. The pipe is visible only along its axis (from the top), but
not at its sides.

O2. The pipe generated in castor oil is less visible from the top
with respect to glycerol and glycerol mixtures.

O3. The pipe generated in pure glycerol becomes more glossy than
the bulk glycerol if illuminated along its axis (from the top)
with normal (not laser) light.

O4. The pipe generated in pure glycerol produces a shadow (with
respect to the bulk glycerol) on a screen when illuminated at its
sides with normal light of not extremely high intensity. (For very
high light intensities, both the pipe and the bulk glycerol appear
on the screen with practically the same luminosity, so that it is
hard to discriminate among them.)

O5. The projection of the pipe (illuminated at its sides) on a
screen shows a greater luminosity (with respect to the bulk
glycerol) at the pipe surface, while its internal part is dark,
even with respect to the bulk glycerol. This can be observed when
the light impinges perpendicularly both on the tube and on the
pipe surface. By increasing the angle of incidence, the brighter
edges increase in width ``invading'' the dark internal part, till
they completely cover the internal part. In such a case, all the
projection of the pipe is brighter with respect to the bulk
glycerol.

O6. The projection of the pipe generated in castor oil is barely
visible on a screen, when illuminated at its sides with normal
light, since its ``shadow'' on the screen is slightly brighter
than the bulk liquid.

O7. The bright/dark effect effect of the point O5, including its
dependence on the angle of incidence, is not observed in the pipes
formed in castor oil.

O8. A laser beam (of sufficiently high intensity), impinging on a
pipe generated in glycerol, is scattered normally to the pipe
surface, the light disc on the screen becoming elliptical in its
shape (see Fig.\,\ref{figO8}). Such scattering is similar to that
from a glass (or plastic) tube filled with a fluid (air, water or
glycerol) and immersed into a tube filled with glycerol, but
remarkable differences arises (see Fig.\,\ref{figO8b}). Indeed, in
such an ``ideal'' case the light disc is scattered to form a
segment which is very thin and long, and extends to both parts of
the glass tube when the laser beam impinges tangentially on it.
Instead, for non-tangential incidence, the scattered beam results
largely deviated on the screen, the deviation being smaller when
filled with glycerol, while greater when filled with water or air.

\begin{figure}
\begin{center}
\begin{tabular}{ccc}
\epsfxsize=3.5cm %
\epsfysize=2.5cm %
\epsffile{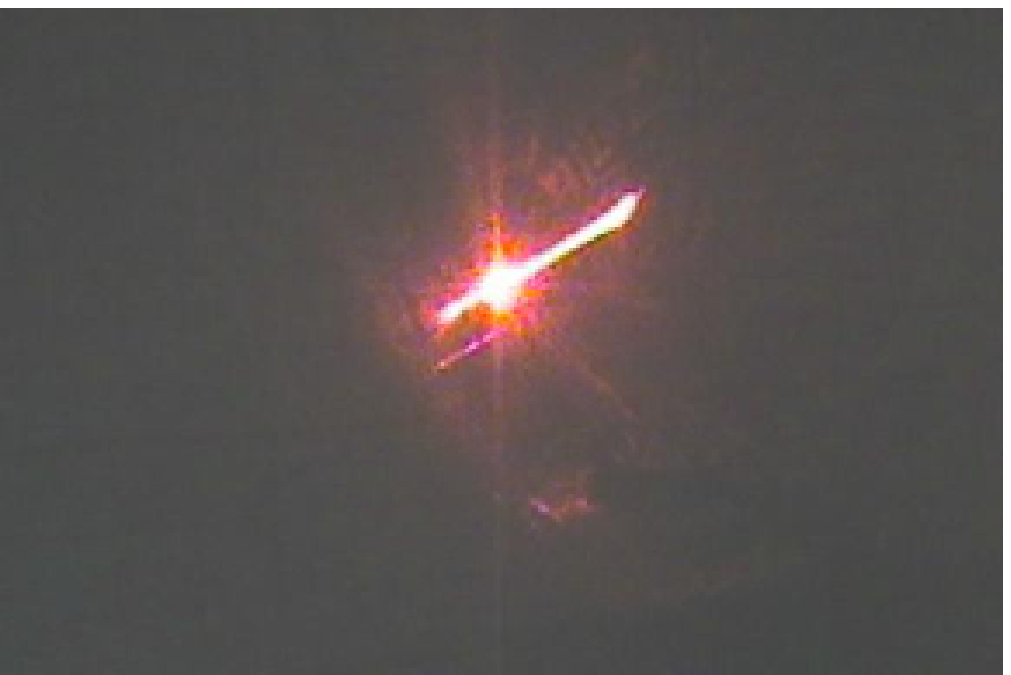} a) & &
\epsfxsize=3.5cm %
\epsfysize=2.5cm %
\epsffile{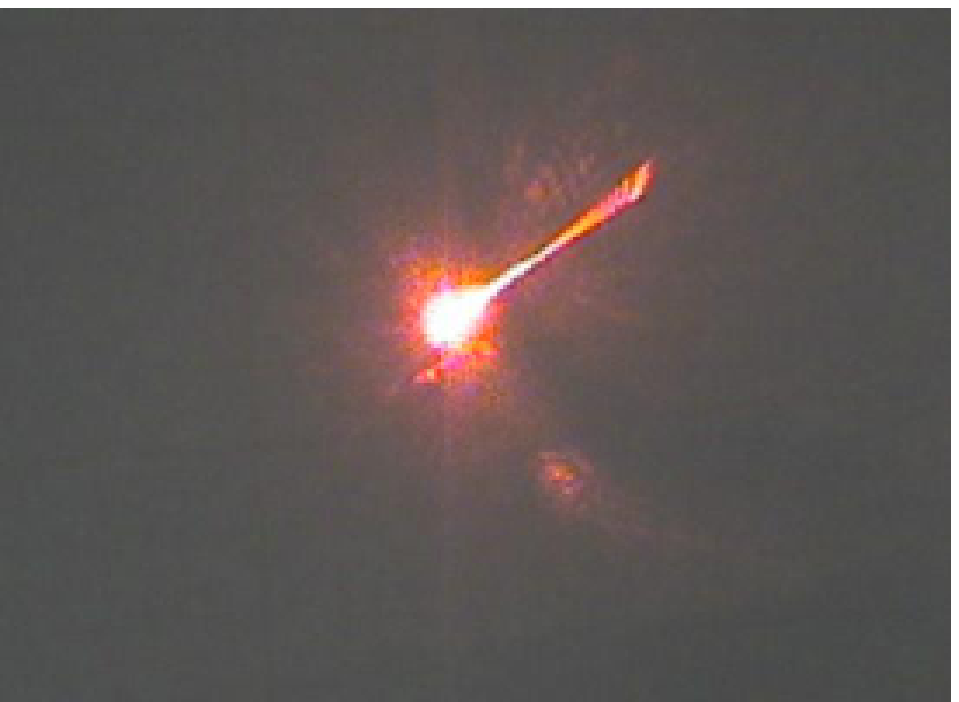} b) \\
\epsfxsize=3.5cm %
\epsfysize=2.5cm %
\epsffile{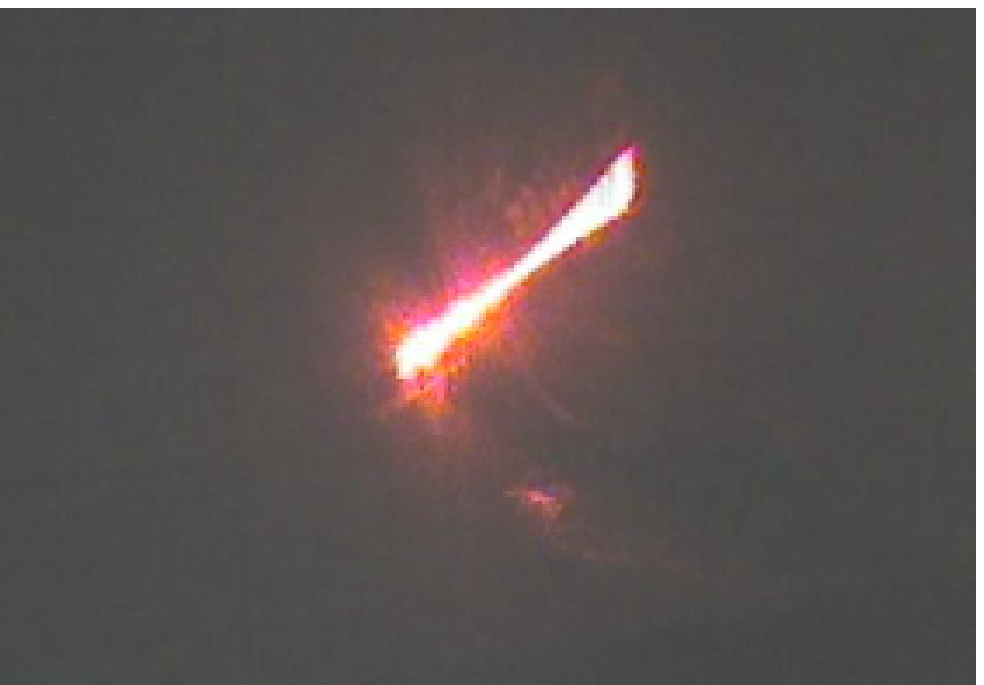} c) & &
\epsfxsize=3.5cm %
\epsfysize=2.5cm %
\epsffile{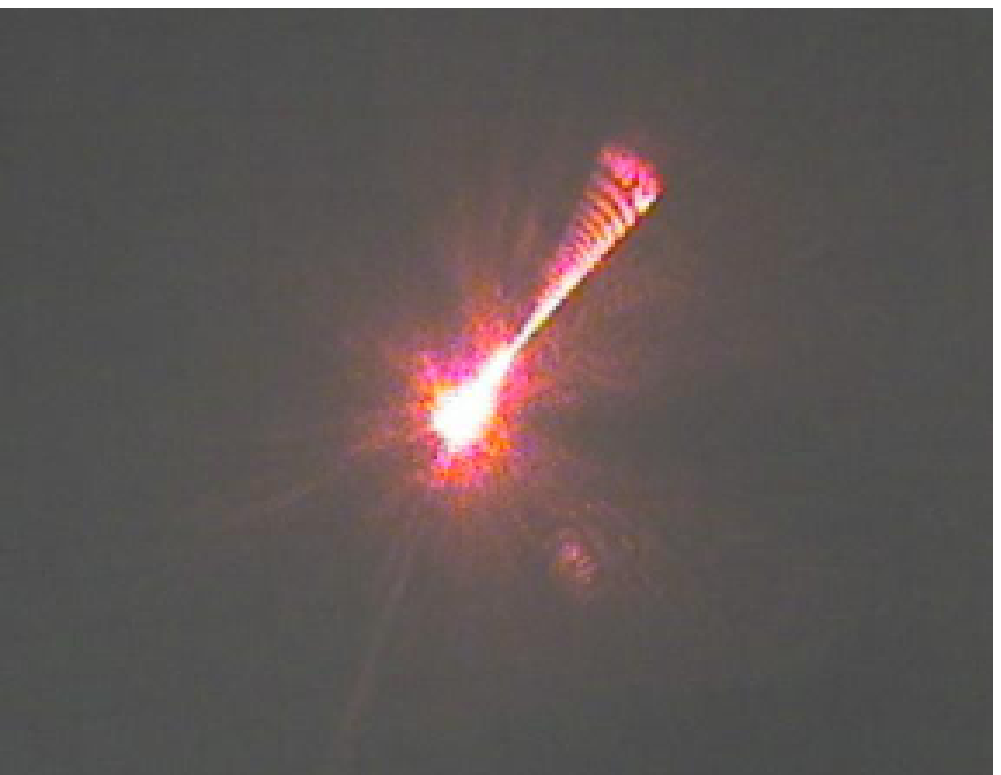} d) \\
\epsfxsize=3.5cm %
\epsfysize=2.5cm %
\epsffile{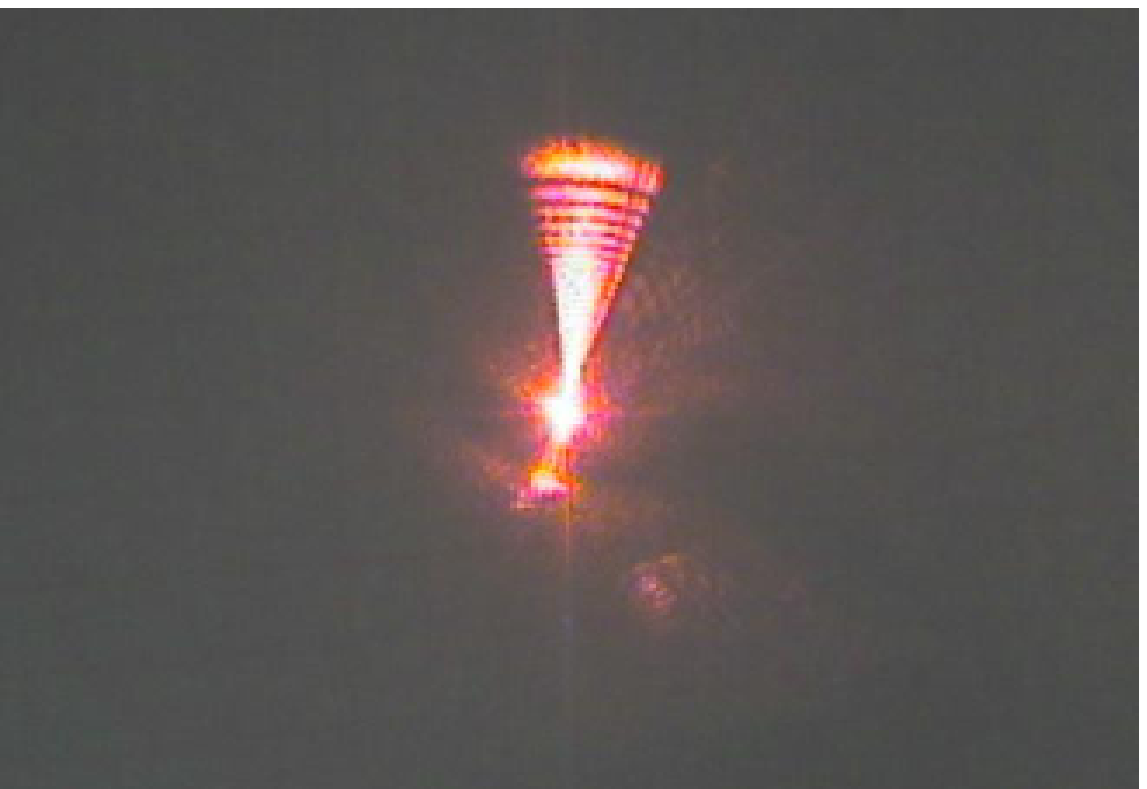} e) & &
\epsfxsize=3.5cm %
\epsfysize=2.5cm %
\epsffile{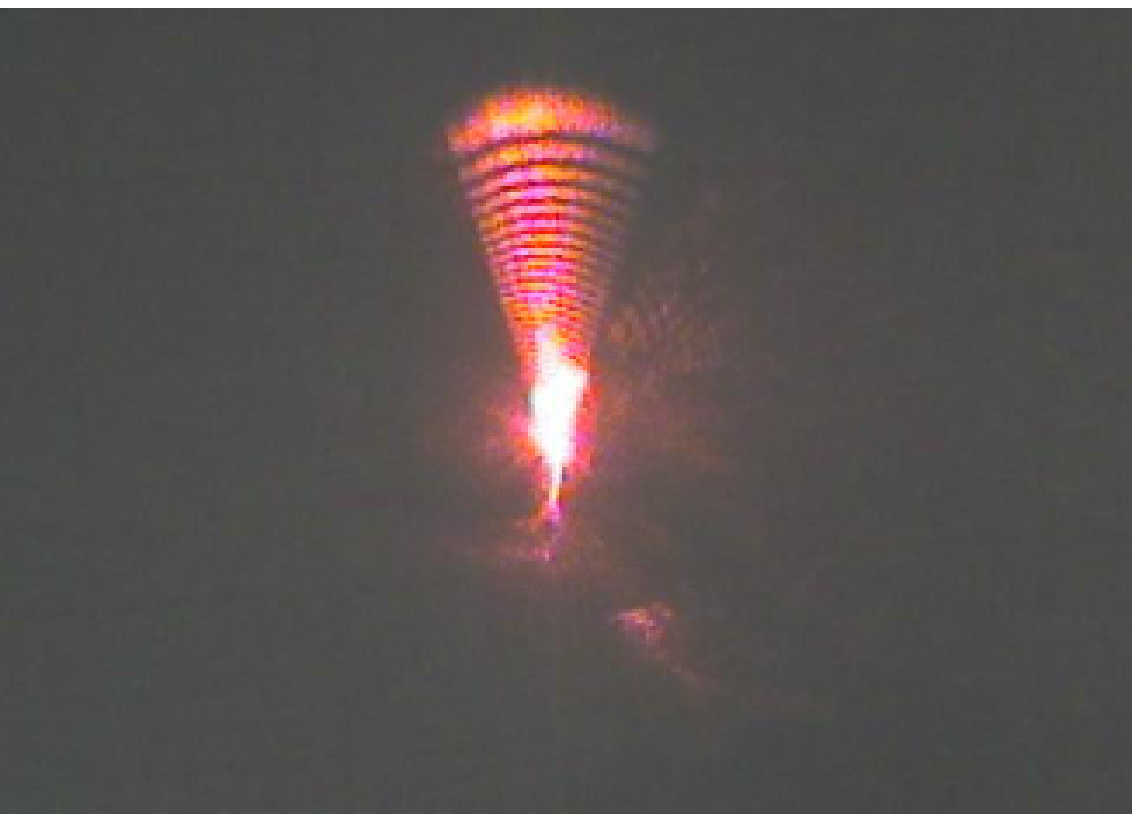} f)
\end{tabular}
\caption{Scattering of a laser beam from an inhomogeneous surface
of a pipe generated in glycerol (O9 effect, see the text).
\newline} \label{figO9}
\end{center}
\vspace{-0.7truecm}
\end{figure}

O9. When the pipe generated in glycerol is not regularly
(cylindrical) shaped, or some inhomogeneities are present near its
surface, the laser beam scattered from such surface presents
(apart from its long and narrow form discussed above) bright
fringes alternated with dark ones. These are visible for any angle
of incidence, even when the laser impinges on the backward lateral
surface of the pipe (see Fig.\,\ref{figO9}).

O10. The O8 effect is observed even for pipes generated in
glycerol/water mixtures, with similar results.

O11. The O8 effect is not observed in pipes generated in castor
oil.

\subsubsection{Secondary effects}

S1. If the steel sphere is left to fall in the glycerol starting
from a point outside it (i.e., in air), microscopic air bubbles
are present in the pipe along its full length. Such micro-bubbles
climb up very slow (they are, practically, trapped inside the
pipe), and render much more visible the pipe when illuminated.

S2. The trapping of air micro-bubbles (S1 effect) is present even
in the pipe generated in castor oil, but the micro-bubbles are
more scattered among them (or less concentrated) with respect to
glycerol.

S3. After some time (parting time) from pipe formation, a fluid
mass parts from the sphere boundary at the bottom of the pipe
formed in glycerol/water mixtures, such mass being previously
``glued'' (due to its high viscosity and surface tension) to the
sphere. The fluid mass climbs up very slowly along the pipe, this
widening out and taking again its original shape after the passage
of the mass (as in the M2 effect).

S4. For a given diameter of the steel sphere, the parting time of
the S3 effect depends on the water concentration of the mixture,
the time being smaller for a greater content of water.

S5. For a given water concentration of the mixture, the parting
time of the S3 effect depends on the diameter of the generating
sphere, the time being smaller for smaller diameters.

\subsubsection{Density and viscosity of the pipe}

P1. The pipes generated in pure glycerol not in a vertical
position, but anyhow inclined, slowly climb up in the bulk
glycerol. This unambiguously shows that the density of the fluid
inside the pipe is slightly lower than that of the bulk glycerol.

P2. In any working condition, the terminal velocity of spheres
falling inside a pipe is slightly greater than that of spheres
falling in the bulk glycerol. This unambiguously shows that the
viscosity of the fluid inside the pipe is slightly lower than that
of the bulk glycerol.

\

Although quite unfavorable experimental conditions have prevented
us to be accurate in the determination of the density and the
viscosity of the pipe liquid, nevertheless we have established
that the per cent variation of such physical quantities has a
rough upper limit of about $4\%$. For example, we have obtained
that: %
\be %
\frac{\eta_{\rm pipe}}{\eta_{\rm bulk}} = 0.96 \pm 0.15
\label{visc} \ee %
and similarly for the density.

\subsubsection{Summarizing remarks}

From the collection of qualitative observations reported above it
comes out quite clearly that the falling of a heavy sphere (of
whatever diameter) induces a modification of the viscous fluid
along its path, resulting in the formation of a definite surface
bounding the pipe. The appearance of such a surface is mainly
responsible of the various mechanical and optical effects reported
above; the physical properties of the liquid used themselves
influence the dynamical evolution of the pipe and the geometrical
properties of the pipe. The development of the phenomenon observed
is certainly related to relaxation processes taking place in the
viscous liquids employed (see, e.g., the point F2), but these
``standard'' processes appear to be not the main source of the
phenomenon itself. Secondary effects, as those mentioned above,
may easily hide the primary phenomenon concerning the ``true''
dynamics of the pipe, so that we have taken particularly care of
avoiding such spurious effects (whenever possible) in our
experimental measurements.

\subsection{Quantitative observations}

\subsubsection{Extinction rates of the pipe}

\begin{table}
\begin{center}
\footnotesize
\begin{tabular}{|r|l|l|l|}
\multicolumn{4}{c}{$x_{\rm glycerol} = 0.99$} \\
\hline
 & $d_0$ & $t_0$ (min) & $\tau$ (min)
\\ \hline
Phase 1 &  $0.202 \pm 0.005$ & $-0.8 \pm 0.3$ & $0.8 \pm 0.2$    \\
\hline
Phase 2 &  $0.174 \pm 0.001$ & $0.8 \pm 0.5$ & $1.9 \pm 0.3$  \\
\hline
Phase 3 &  $0.150 \pm 0.009$ & $1 \pm 8$ & $8\pm 6$    \\
\hline
Phase 4 &  $0.126 \pm 0.006$ & $19 \pm 5$ & $8 \pm 5$     \\
\hline
\multicolumn{4}{c}{$x_{\rm glycerol} = 0.98$} \\
\hline
 & $d_0$ & $t_0$ (min) & $\tau$ (min)
\\ \hline
Phase 1 &  &  &   \\
\hline
Phase 2 &  $0.155 \pm 0.003$ & $-0.4 \pm 0.1$ & $0.9 \pm 0.1$    \\
\hline
Phase 3 &  $0.1398 \pm 0.0003$ & $4.7 \pm 0.4$ & $1.4 \pm 0.2$  \\
\hline
Phase 4 &  $0.128 \pm 0.001$ & $14 \pm 6$ & $2\pm 2$    \\
\hline
\multicolumn{4}{c}{$x_{\rm glycerol} = 0.97$} \\
\hline
 & $d_0$ & $t_0$ (min) & $\tau$ (min)
\\ \hline
Phase 1 &  &  &   \\
\hline
Phase 2 &  &  &   \\
\hline
Phase 3 &  $0.144 \pm 0.002$ & $0.23 \pm 0.07$ & $0.61 \pm 0.9$    \\
\hline
Phase 4 &  $0.11 \pm 0.01$ & $6 \pm 5$ & $8 \pm 6$  \\
\hline
\multicolumn{4}{c}{$x_{\rm glycerol} = 0.96$} \\
\hline
 & $d_0$ & $t_0$ (min) & $\tau$ (min)
\\ \hline
Phase 1 &  &  &   \\
\hline
Phase 2 &  $0.17 \pm 0.01$ & $-1.1 \pm 0.9$ & $0.9 \pm 0.7$    \\
\hline
Phase 3 &  $0.1219 \pm 0.0007$ & $1.0 \pm 0.1$ & $2.2 \pm 0.1$  \\
\hline
Phase 4 &  $0.112 \pm 0.004$ & $0.3 \pm 8.9$ & $4 \pm 4$  \\
\hline
\multicolumn{4}{c}{$x_{\rm glycerol} = 0.95$} \\
\hline
 & $d_0$ & $t_0$ (min) & $\tau$ (min)
\\ \hline
Phase 1 &  &  &   \\
\hline
Phase 2 &  $0.172 \pm 0.001$ & $0.4 \pm 0.1$ & $0.9 \pm 0.1$    \\
\hline
Phase 3 &  $0.15 \pm 0.01$ & $1 \pm 3$ & $2 \pm 2$  \\
\hline
Phase 4 &  $0.129 \pm 0.002$ & $3 \pm 2$ & $4 \pm 1$  \\
\hline
\end{tabular}
\caption{Fitting parameters for the extinction rates in Eq.\,(\ref{eqdratio}) (see the text). \newline} \label{rates}
\end{center}
\vspace{-0.7truecm}
\end{table}
\setcounter{table}{1}
\begin{table}
\begin{center}
\footnotesize
\begin{tabular}{|r|l|l|l|}
\multicolumn{4}{c}{$x_{\rm glycerol} = 0.94$} \\
\hline
 & $d_0$ & $t_0$ (min) & $\tau$ (min)
\\ \hline
Phase 1 &  &  &   \\
\hline
Phase 2 &  &  &   \\
\hline
Phase 3 &  $0.156 \pm 0.003$ & $0.4 \pm 0.4$ & $2.8 \pm 0.5$    \\
\hline
Phase 4 &  $0.1465 \pm 0.0002$ & $6.7 \pm 0.3$ & $2.2 \pm 0.1$  \\
\hline
\multicolumn{4}{c}{$x_{\rm glycerol} = 0.93$} \\
\hline
 & $d_0$ & $t_0$ (min) & $\tau$ (min)
\\ \hline
Phase 1 &  &  &   \\
\hline
Phase 2 &  &  &   \\
\hline
Phase 3 &  $0.149 \pm 0.005$ & $0.9 \pm 0.4$ & $2.3 \pm 0.5$    \\
\hline
Phase 4 &  $0.111 \pm 0.003$ & $6.4 \pm 0.9$ & $4 \pm 1$  \\
\hline
\multicolumn{4}{c}{$x_{\rm glycerol} = 0.92$} \\
\hline
 & $d_0$ & $t_0$ (min) & $\tau$ (min)
\\ \hline
Phase 1 &  &  &   \\
\hline
Phase 2 &  &  &   \\
\hline
Phase 3 &  $0.149 \pm 0.004$ & $1.3 \pm 0.2$ & $2.9 \pm 0.2$    \\
\hline
Phase 4 &  $0.106 \pm 0.005$ & $11.2 \pm 0.8$ & $5 \pm 1$  \\
\hline
\multicolumn{4}{c}{$x_{\rm glycerol} = 0.91$} \\
\hline
 & $d_0$ & $t_0$ (min) & $\tau$ (min)
\\ \hline
Phase 1 &  &  &   \\
\hline
Phase 2 &  &  &   \\
\hline
Phase 3 &  $0.1105 \pm 0.0008$ & $1.06 \pm 0.03$ & $0.97 \pm 0.05$    \\
\hline
Phase 4 &  &  &   \\
\hline
\multicolumn{4}{c}{$x_{\rm glycerol} = 0.90$} \\
\hline
 & $d_0$ & $t_0$ (min) & $\tau$ (min)
\\ \hline
Phase 1 &  &  &   \\
\hline
Phase 2 &  &  &   \\
\hline
Phase 3 &  &  &   \\
\hline
Phase 4 &  $0.108 \pm 0.003$ & $1.48 \pm 0.09$ & $0.9 \pm 0.2$    \\
\hline
\end{tabular}
\caption{Fitting parameters for the extinction rates in Eq.\,(\ref{eqdratio}) (cont.). \newline} \label{tabledratio}
\end{center}
\vspace{-0.7truecm}
\end{table}

\noi Measurements of the diameter of the pipe divided by that of
the falling sphere ($d_{\rm ratio} = d_{\rm pipe}/d_{\rm sphere}$)
at given instants of time since pipe formation (chosen as the
reference time $t=0$) have been performed as a function of the
concentration of glycerol/water mixtures. Such observations have
been carried out by illuminating the sample with normal (non
laser) light and measuring the size of the shadow of the pipe and
the sphere projected laterally on a screen. The given measurement
ended when an appreciable shadow (with respect to the bulk liquid)
was no more projected on the screen; note, however, that at this
time the pipe continued to persist, as it has been observed by
looking directly at the pipe from its top (see the point O1). The
following measurements have been performed with spheres of
diameter 5 mm, at a temperature of $24 \div 25 ^{\rm o}$C.

\

\begin{figure}
\begin{center}
\begin{tabular}{ccc}
\epsfxsize=3.6truecm %
\epsffile{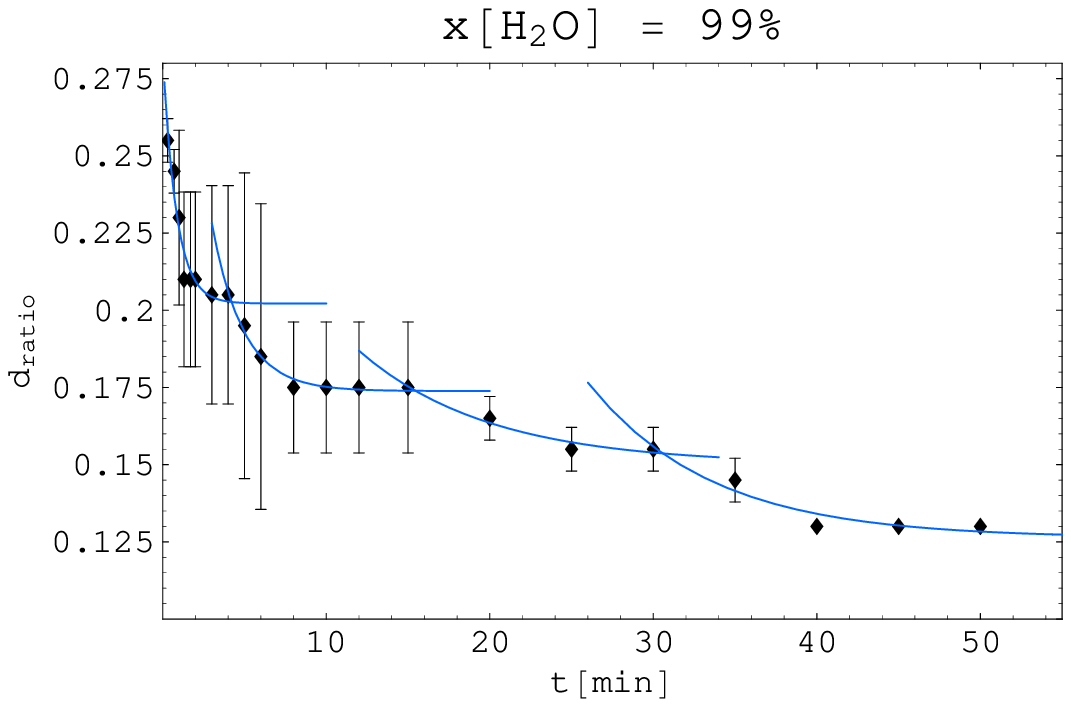} a) & &
\epsfxsize=3.6truecm %
\epsffile{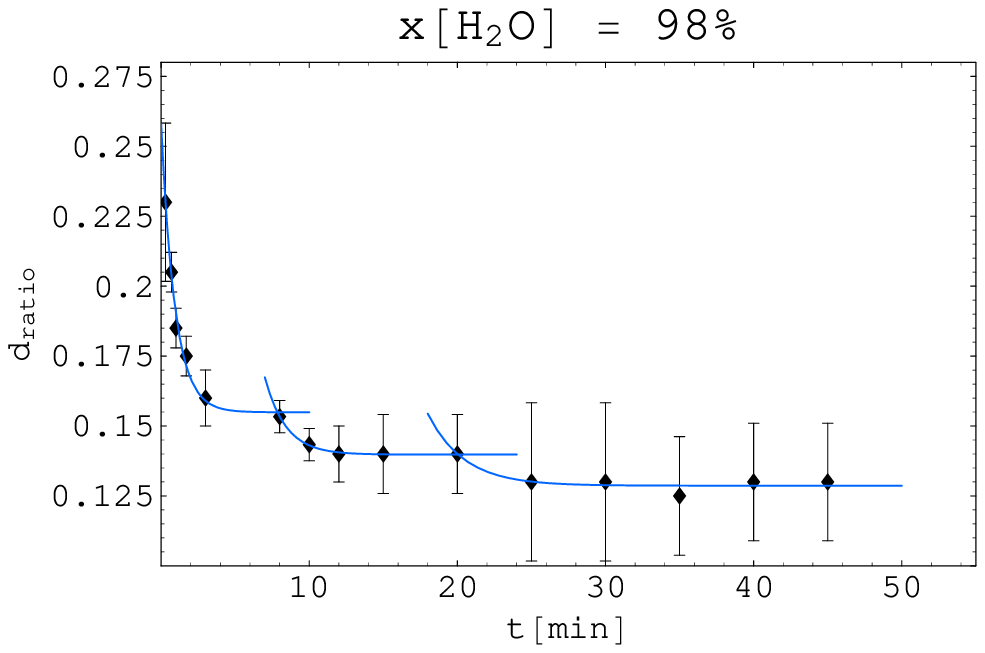} b) \\
\epsfxsize=3.6truecm %
\epsffile{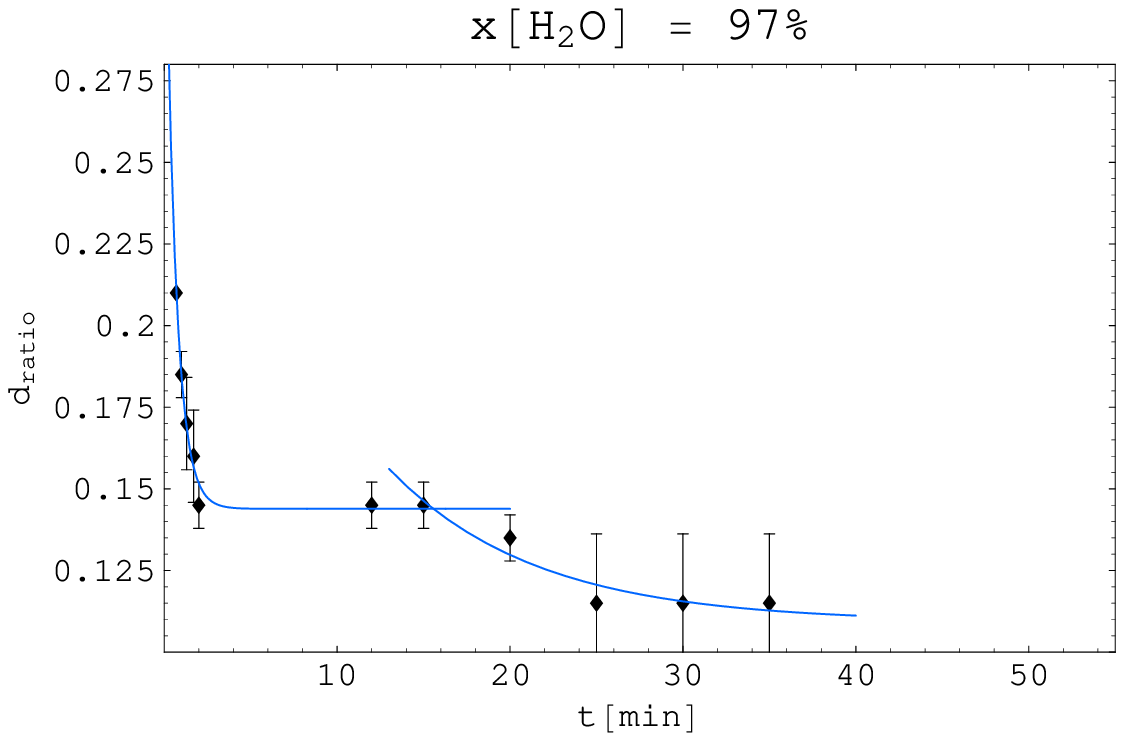} c) & &
\epsfxsize=3.6truecm %
\epsffile{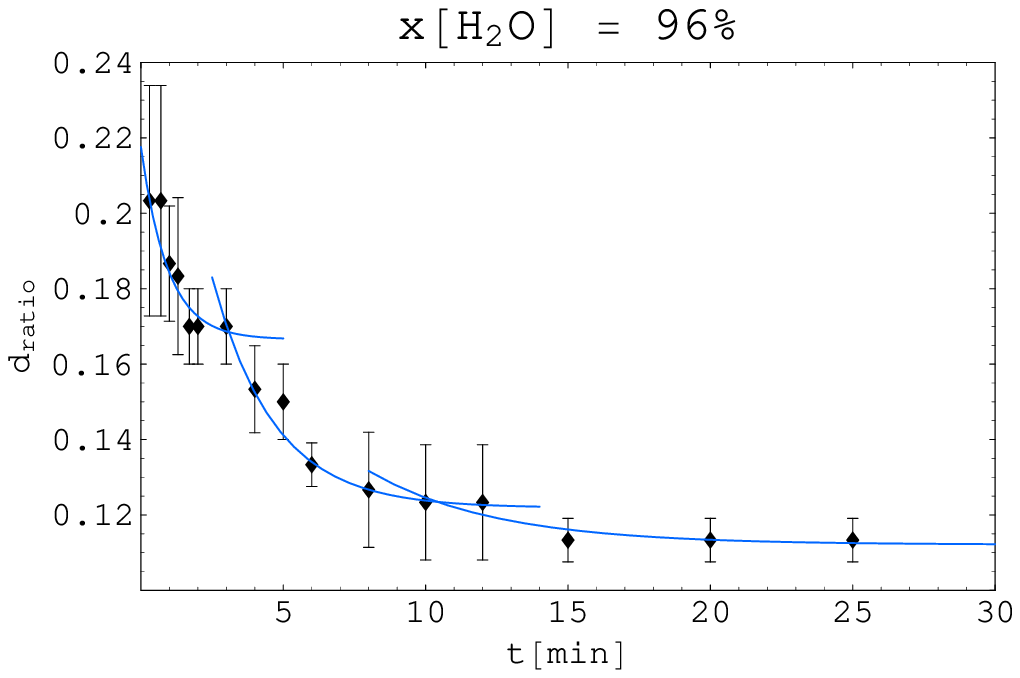} d) \\
\epsfxsize=3.6truecm %
\epsffile{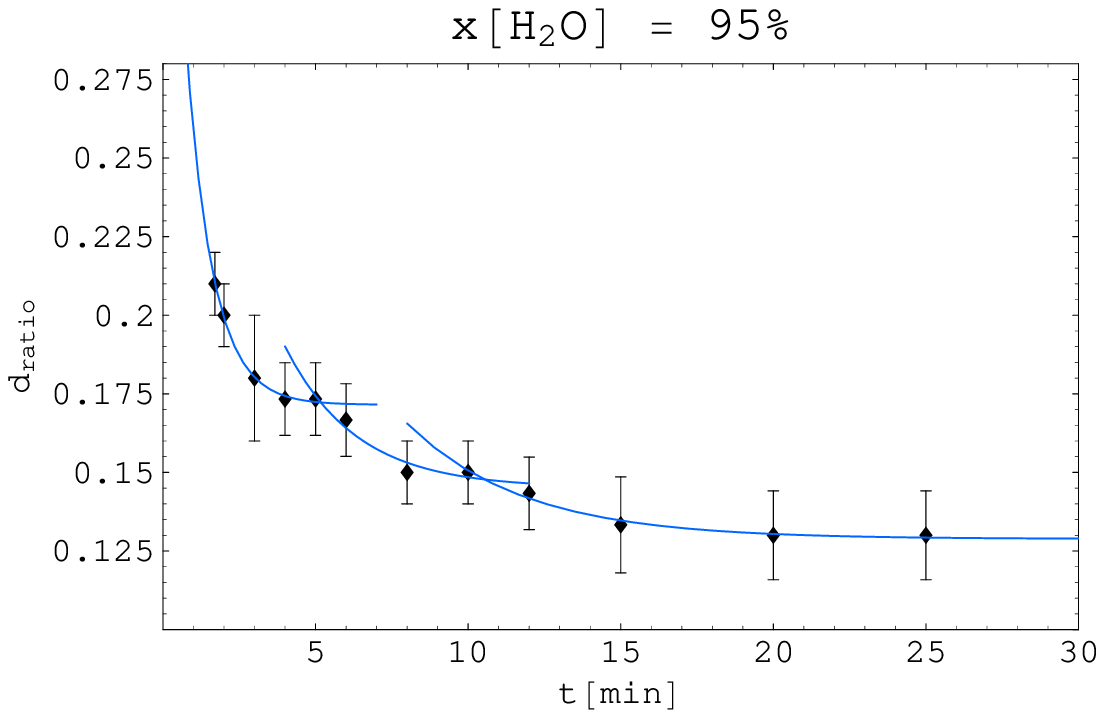} e) & &
\epsfxsize=3.6truecm %
\epsffile{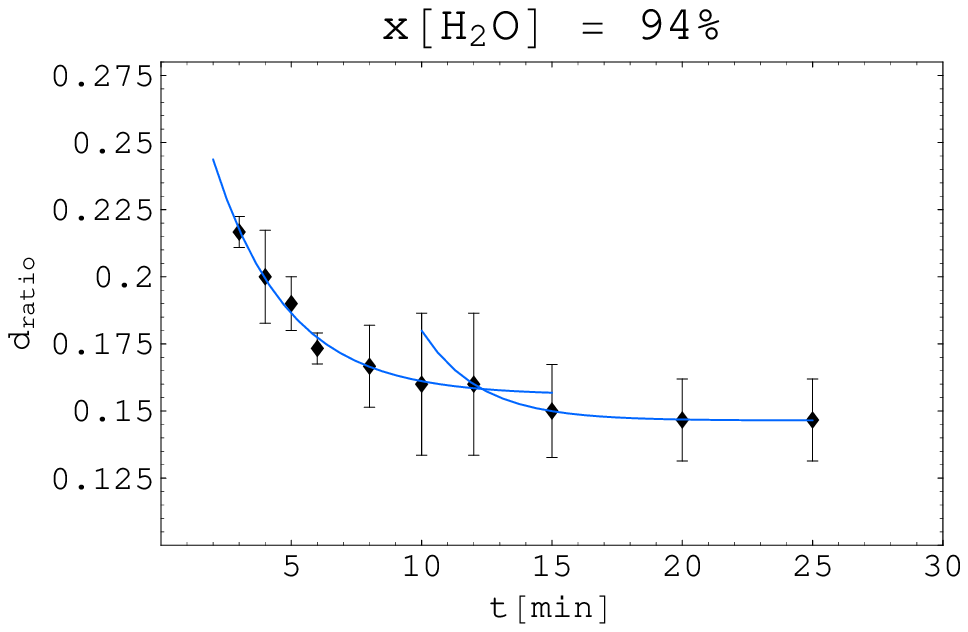} f) \\
\epsfxsize=3.6truecm %
\epsffile{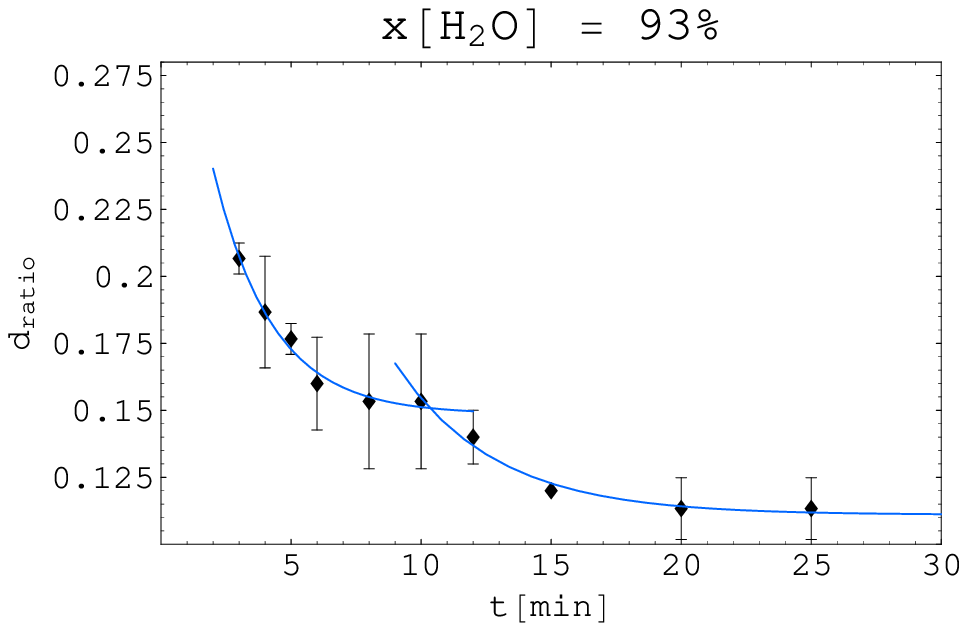} g) & &
\epsfxsize=3.6truecm %
\epsffile{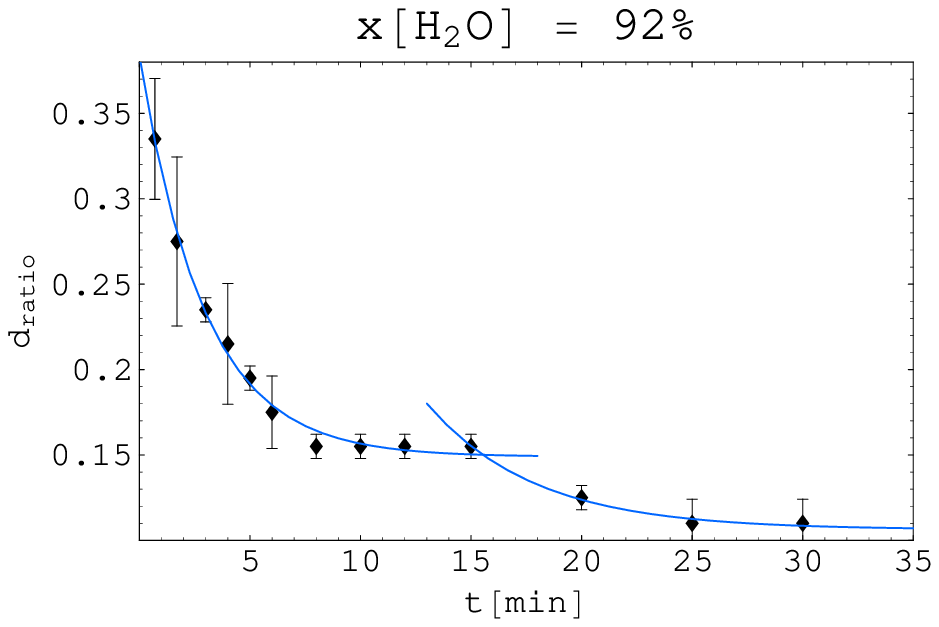} h) \\
\epsfxsize=3.6truecm %
\epsffile{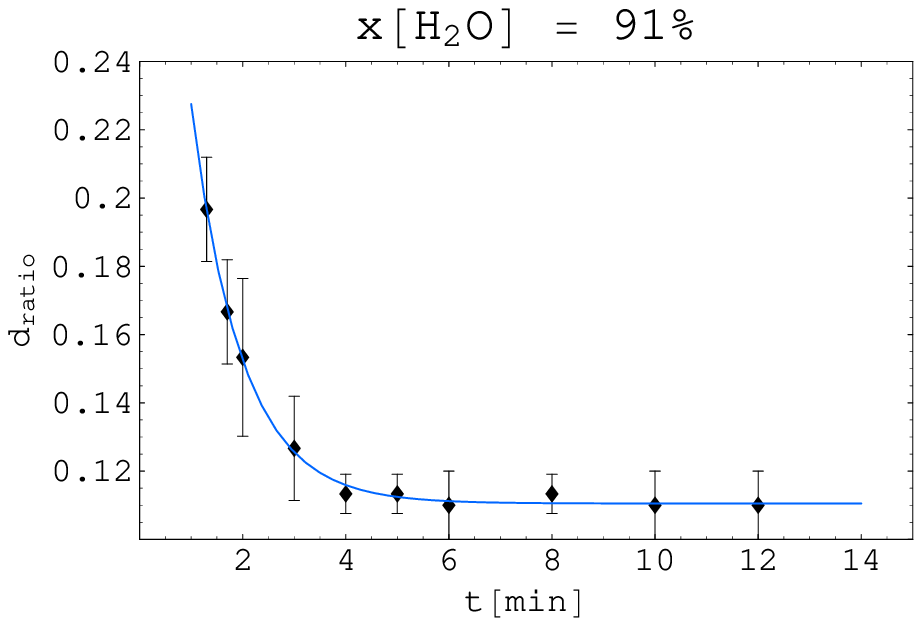} i) & &
\epsfxsize=3.6truecm %
\epsffile{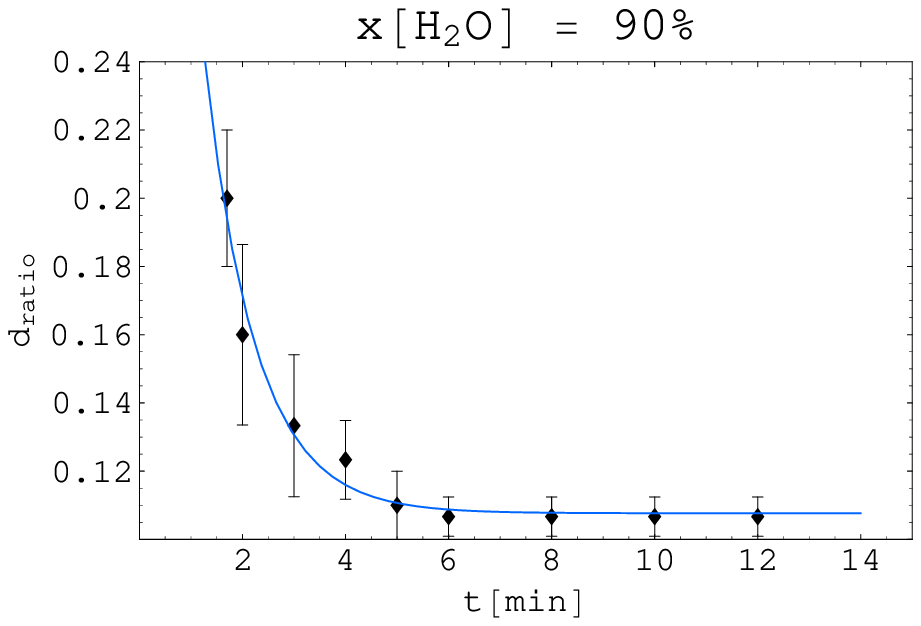} l)
\end{tabular}
\caption{Time evolution of the pipe radius for different water
contents of the glycerol/water mixture (see the text). \newline}
\label{figdratio}
\vspace*{-1.4truecm}
\end{center}
\end{figure}

R1. The time evolution of the pipe radius has always an
exponentially decreasing behavior (in given regions of time), for
any concentration of the glycerol/water mixture. The fitting
function employed with the set of
measurements obtained is the following: %
\be \label{eqdratio} %
d_{\rm ratio} = d_0 \left( \erm^{-\frac{t - t_0}{\tau}} + 1
\right) .
\ee %
The values of the fitting parameters are reported in Table
\ref{tabledratio}; note that the assignment of a certain parameter
to a given ``phase'' (see the point R2) is somewhat arbitrary, and
(partially) justified only {\it a posteriori}, by comparing the
different values for different concentrations (and also
introducing some theoretical bias; see Section III).
In Fig.\,\ref{figdratio} we show the experimental data together with the
fitting curve from Eq.\,(\ref{eqdratio}).

\

The experimental points refer to averages over several sets of
measurements, performed at given concentrations; the big error
bars occurring in some cases are due to the (relatively) big
deviation between these sets of data. Although the fitting
procedures have been carried out on the average values obtained,
we have always complied with the general information contained in
{\em each} set of data and agreeing among those pieces of
information. Thus, for example, although the big errors apparently
do not support completely the adoption of a given fitting function
rather than another, such adoption comes out, instead, from the
statistical analysis of single sets of data. The same reasoning
applies to the choice of the starting and the ending of the
different fitting regions: the experimental data in two (or more)
adjacent regions that, looking at average values, may be fitted by
a single function within the errors, have been described by two
(or more) different fitting curves, in agreement with what shown
by any of the single sets of data conveying in the showed average
values.

R2. The dynamical evolution of the pipe generated in
glycerol/water mixtures takes place in different, subsequent time
phases, each one characterized by its own exponential behavior.
The trigger of such phases depends on the concentration of the
mixture; the trigger times of the different phases (roughly, the
intersection points of the different exponential curves, for given
concentration) are reported in Table \ref{tabletrigger}.

\subsubsection{Persistence time of the pipe}

\noi The persistence time $t_{\rm persist}$ of the pipe is defined
as the time occurred since pipe formation till its visible
disappearance, that is, till the intensity of the light reflected
by the pipe and revealed from the top (parallel to the pipe axis)
reduces below a given experimental threshold. We have performed a
number of different runs aimed at measuring this quantity for
pipes generated in different liquids, as function of key
parameters (concentration and temperature), the results obtained
being reported and discussed below.

\

\begin{figure}
\begin{center}
\epsfxsize=8cm %
\epsffile{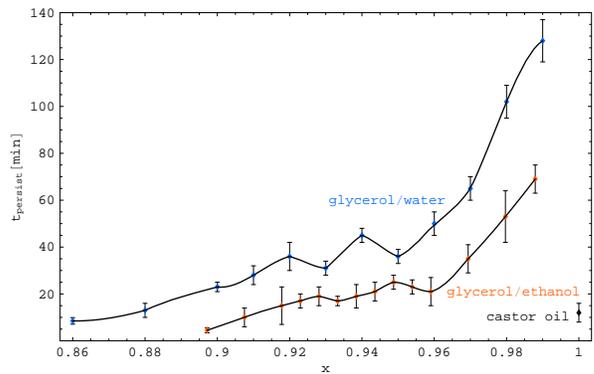} \caption{Persistence time of pipes
formed in glycerol/water ($24 ^{\rm o}$C) and glycerol/ethanol
($24.5 ^{\rm o}$C) mixtures as function of the glycerol
concentration. For reference, the experimental point for the
persistence time in pure castor oil ($23 ^{\rm o}$C) is reported
as well (the dependence on the concentration for castor oil
mixtures is practically unobservable with the presently adopted
apparatus, due to the extremely low value of the persistence
time).
\newline} \label{figwatethancast}
\end{center}
\vspace{-0.7truecm}
\end{figure}

R3. The pipe persistence time broadly decreases with decreasing
glycerol concentration in glycerol mixtures reaching, however, two
relative maxima for values of the concentration approximately
given by:
\be \label{eqmaxw} %
x_{\rm glycerol} \simeq 0.92, \qquad \qquad x_{\rm glycerol}
\simeq 0.94
\ee %
for glycerol/water and:
\be \label{eqmaxe} %
x_{\rm glycerol} \simeq 0.93, \qquad \qquad x_{\rm glycerol}
\simeq 0.95
\ee %
for glycerol/ethanol mixtures, respectively. The experimental
points are reported in Fig.\,\ref{figwatethancast} (we have used 3
mm spheres, at a temperature of $24 ^{\rm o}$C and $24.5 ^{\rm
o}$C for glycerol/water and glycerol/ethanol mixtures,
respectively).

R4. In the evolution dynamics of the pipe for\-med in
glycerol/water and glycerol/ethanol mixtures, three different
contributions have been found in the $t_{\rm persist}(x_{\rm
glycerol})$ plot; they are quite well described by gaussian
fitting curves (see Fig.\,\ref{figfitpersist}):
\be \label{eqpersx} %
t_{\rm persist} = t_a \erm^{- \frac{(x_{\rm glycerol} -
x_0)^2}{x_\sigma^2}} + t_b,
\ee %
where the fitting parameters are reported in Table
\ref{tablepers}. Note that, in the region of greater concentration
(case 1), the parameter $x_0$ has been fixed to the maximum
content of glycerol in the mixture. Instead, in the intermediate
region of concentration (case 2), at least for glycerol/water
mixtures, the reported values of the fitting parameters are purely
indicative, since the small extension of the probed region does
not allow an accurate determination of the parameters of the
gaussian curve (near its maximum).

\begin{table}
\begin{center}
\footnotesize
\begin{tabular}{|r|l|l|l|}
\hline
 & $t_{1 \rightarrow 2}$ (min)
 & $t_{2 \rightarrow 3}$ (min)
 & $t_{3 \rightarrow 4}$ (min)
\\ \hline
$x_{\rm glycerol} = 0.99$ & 4.2 & 15.5 & 30.6 \\
\hline
$x_{\rm glycerol} = 0.98$ & & 7.8 & 20.0  \\
\hline
$x_{\rm glycerol} = 0.97$ & & & 15.6   \\
\hline
$x_{\rm glycerol} = 0.96$ & & 3.1 & 10.4  \\
\hline
$x_{\rm glycerol} = 0.95$ & & 5.2 & 10.5  \\
\hline
$x_{\rm glycerol} = 0.94$ & & & 12.3   \\
\hline
$x_{\rm glycerol} = 0.93$ & & & 10.3   \\
\hline
$x_{\rm glycerol} = 0.92$ & & & 15.5   \\
\hline
$x_{\rm glycerol} = 0.91$ & & & --   \\
\hline
$x_{\rm glycerol} = 0.90$ & & & --   \\
\hline
\end{tabular}
\caption{Trigger times for the different exponential phases in the
dynamical evolution of the pipe generated in different
glycerol/water mixtures. \newline} \label{tabletrigger}
\end{center}
\vspace{-0.7truecm}
\end{table}

\begin{figure}
\begin{center}
\epsfxsize=8cm %
\epsffile{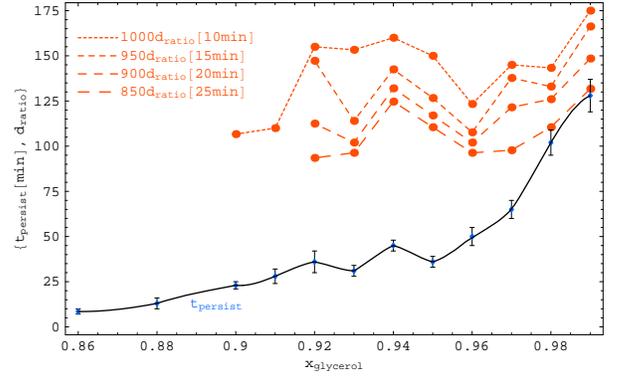} \caption{Comparison between the persistence
time curve (for pipes generated in glycerol/water mixtures) and
the pipe diameters curves at subsequent time instants since
formation, as a function of the glycerol concentration. The
quantities $d_{\rm ratio}$ have been multiplied by a suitable
rescaling factor (reported in the plot), in order to be easily
recognizable, while arbitrary units have been adopted on the
vertical axis. \newline } \label{figpersdratio}
\end{center}
\vspace{-0.7truecm}
\end{figure}

R5. The different contributions in the evolution dynamics of the
pipe (see the point R4) do not manifest only at later times, in
the persistence time curves, but are {\it independently} confirmed
by the evolution curves of the pipe diameter at different times
since formation, as showed in Fig.\,\ref{figpersdratio}.

\begin{figure}
\begin{center}
\begin{tabular}{ccc}
\epsfxsize=3.5cm %
\epsfysize=3cm %
\epsffile{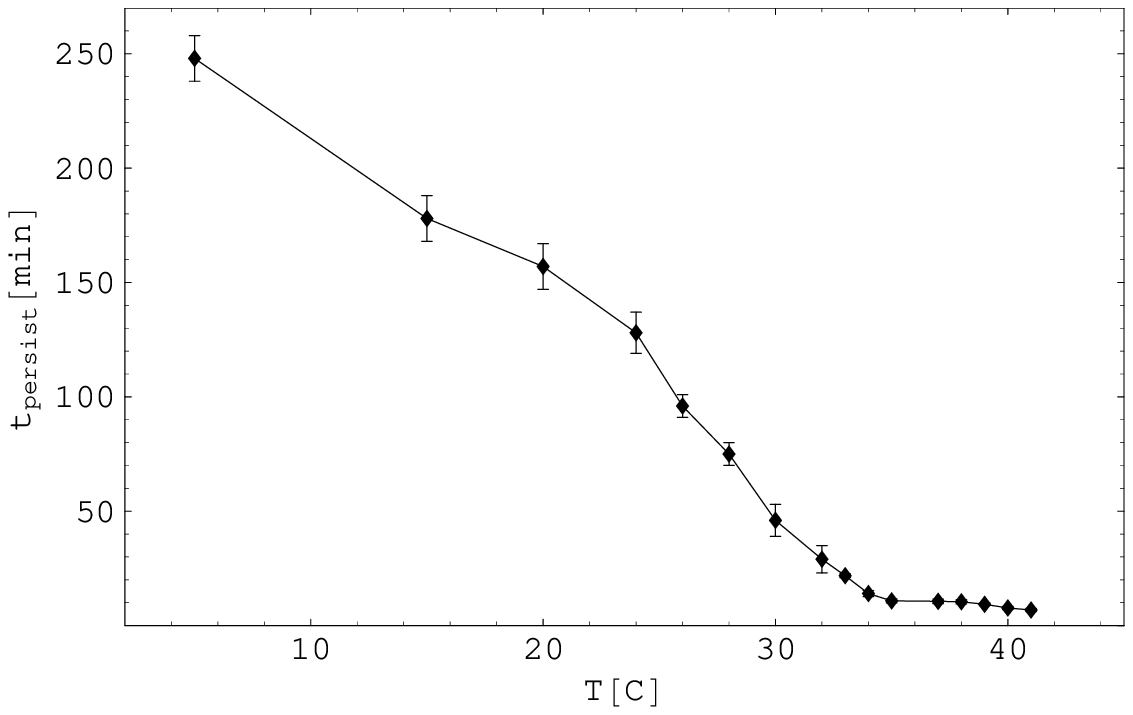} a) & &
\epsfxsize=3.5cm %
\epsfysize=3cm %
\epsffile{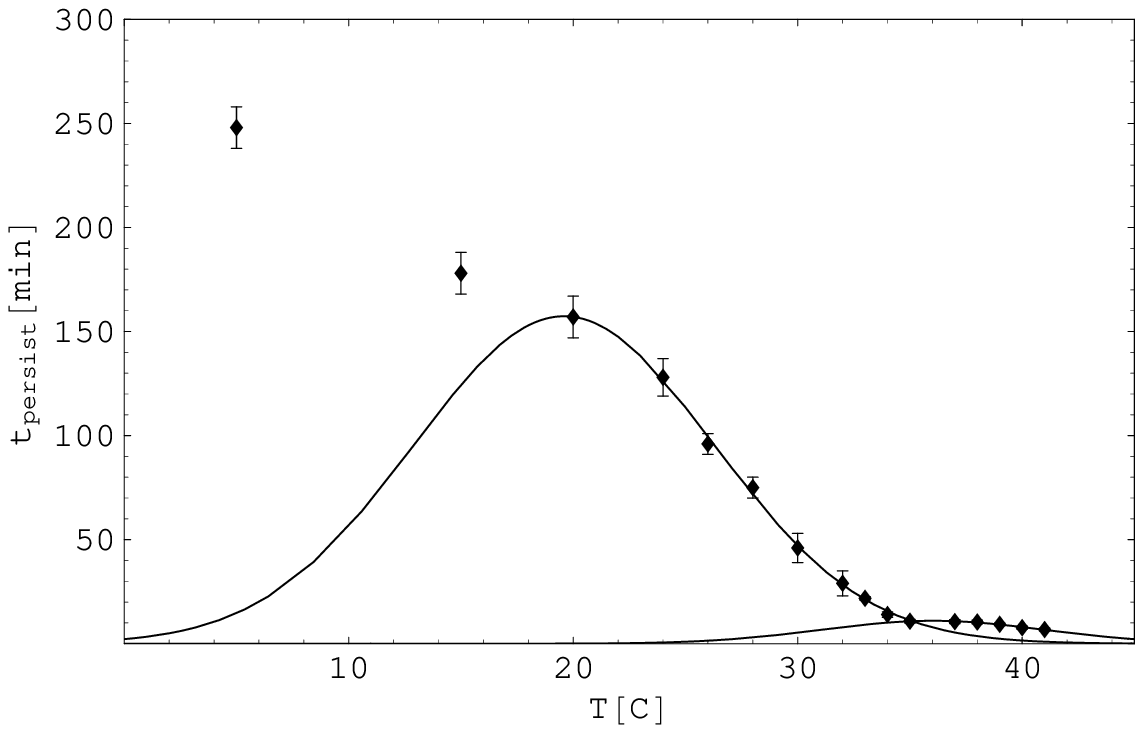} b)
\end{tabular}
\caption{Persistence time for pipes generated in pure glycerol as
function of the temperature: a) interpolation of the experimental
points; b) data fitting with two different gaussian curves.
\newline} \label{figperstemp}
\end{center}
\vspace{-0.7truecm}
\end{figure}

R6. The pipe persistence time (in pure glycerol) decreases with
increasing temperature, according to three different behaviors
(see Fig.\,\ref{figperstemp}a).

R7. In the evolution dynamics of the pipe generated in pure
glycerol, different contributions have been found in the $t_{\rm
persist}(T)$ plot; in the temperature regions where enough data
are available, they are quite well described by gaussian fitting
curves (see Fig.\,\ref{figperstemp}b):
\be \label{eqperst} %
t_{\rm persist} = t_c \erm^{- \frac{(T - T_0)^2}{T_\sigma^2}},
\ee %
where the fitting parameters are reported in Table
\ref{tableperstemp}.

\begin{figure}
\begin{center}
\epsfxsize=8cm %
\epsffile{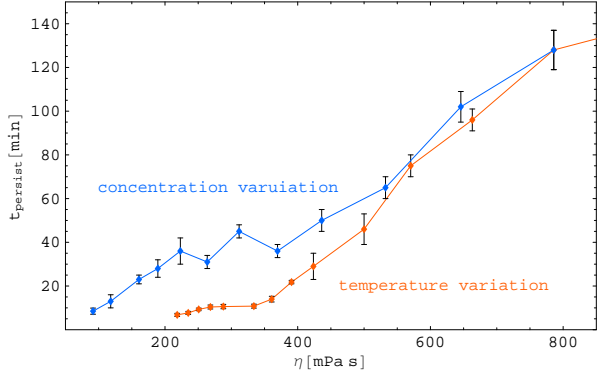} %
\caption{Persistence time curves, for pipes generated in
glycerol/water mixtures, as functions of the viscosity. The
variation of the viscosity has been induced in two different ways:
by varying the glycerol concentration (upper curve) or the
temperature (lower curve) of the mixture. \newline}
\label{figvisc}
\end{center}
\vspace{-0.7truecm}
\end{figure}

R8. The curves $t_{\rm persist}(x_{\rm glycerol})$ and $t_{\rm
persist}(T)$, obtained with evidently different experimental
methods, may be compared when one assumes that the glycerol
concentration and temperature variations contribute essentially to
change only the viscosity $\eta$ of the liquid. In such an
assumption, by using known data on the dependence of $\eta$ on the
concentration and temperature of the mixtures \cite{glycerolprop},
the $t_{\rm persist}(\eta)$ curves reported in Fig.\,\ref{figvisc}
are obtained. From these curves it is evident that the appearance
of different contributions to the evolution dynamics of the pipe
is effectively ruled by the viscosity parameter, though a non
negligible role is nevertheless played directly by the water (or
ethanol) content of the mixture (through the concentration $x_{\rm
glycerol}$) and by the thermal agitation (through the temperature
$T$) for viscosity values lower than approximately 550 mPa
$\!\cdot\!$ s.

R9. The comparison of the persistence time curves for pipes
generated in different viscous liquids reveals a similar behavior
for the phenomenon studied which, however, develops on different
time scales, as shown in Fig.\,\ref{figwatethancast}. Note that the
(extrapolated) matching at $x_{\rm glycerol} =1$ of the curves,
corresponding to glycerol/water and glycerol/ethanol mixtures, is
not smooth.

\begin{figure}
\begin{center}
\epsfxsize=8cm %
\epsffile{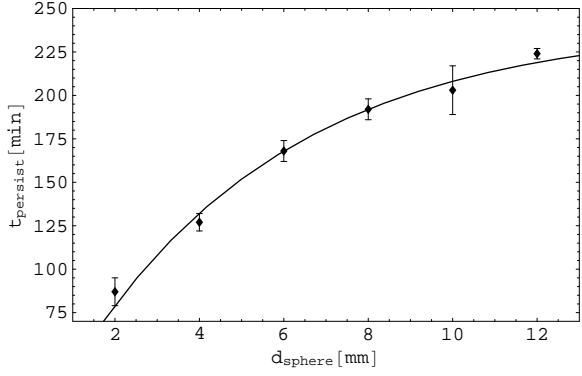} \caption{Persistence time of pipes generated
in pure glycerol ($24.5 \div 25 ^{\rm o}$C) as a function of the
diameter of the generating spheres. \newline} \label{figpersdiam}
\end{center}
\vspace{-0.7truecm}
\end{figure}

R10. The persistence time of pipes generated in pure glycerol
increases exponentially with the diameter of the generating
sphere, as shown in Fig.\,\ref{figpersdiam}. The fitting function
appearing in this plot is:
\be \label{eqpersd} %
t_{\rm persist} = t_d \left( 1 - \erm^{- \frac{d_{\rm sphere}}{D}}
\right),
\ee %
where $t_d = 240 \pm 10$ min and $D = 5.0 \pm 0.5$ mm.

\begin{table}
\begin{center}
\footnotesize
\begin{tabular}{|r|l|l|l|l|}
\multicolumn{5}{c}{Glycerol/water mixture} \\
\hline
 & $t_a$ (min) & $t_b$ (min) & $x_0$ & $10^{4} x_\sigma^2$
\\ \hline
1. & $110 \pm 6$ & $31 \pm 6$ & 1 & $8 \pm 1$ \\
\hline
2. & $17$ & $27$ & $0.941$ & $1.0$ \\
\hline
3.  & $25 \pm 3$ & $9 \pm 2$ & $0.922 \pm 0.003$ & $8 \pm 4$ \\
\hline \multicolumn{5}{c}{Glycerol/ethanol mixture} \\
\hline
 & $t_a$ (min) & $t_b$ (min) & $x_0$ & $10^{4} x_\sigma^2$
\\ \hline
1. & $68 \pm 2$ & $11 \pm 3$ & 1 & $9 \pm 1$ \\
\hline
2. & $17 \pm 4$ & $1 \pm 4$ & $0.929 \pm 0.002$ & $7 \pm 4$ \\
\hline
3.  & $6 \pm 2$ & $19 \pm 2$ & $0.950 \pm 0.001$ & $0.5 \pm 0.4$ \\
\hline
\end{tabular}
\caption{Fitting parameters for the persistence time curves in Eq.\,(\ref{eqpersx}) (see the text). \newline} \label{tablepers}
\end{center}
\vspace{-0.7truecm}
\end{table}

\subsubsection{Summarizing remarks}

\noi The evolution dynamics of the pipes generated in viscous
liquids comes out to be quite similar for different substances,
but takes place on peculiar time scales and tends to be very fast
for not high values of the viscosity, so that for moderately
viscous or inviscid liquids the effect is practically
unobservable.

The phenomenon appears to be completely settled by the liquid
viscosity for $\eta \geq 550$ mPa $\!\cdot\!$ s, and, in this region,
the persistence time increases monotonically with increasing
viscosity. Instead, for lower values of the viscosity, the
microscopic structure of the liquid and its thermal energy play a
non-negligible role: different contributions to the evolution
dynamics come out at any instant of time since pipe formation, and
the persistence time curves exhibit two maxima for peculiar values
of the concentration (depending on the type of mixture used). In
any of these regions, the persistence time curves are quite well
approximated by gaussian fits, irrespective of the particular
liquid mixture employed. It is particularly remarkable the fact
that the appearance of such diverse contributions is confirmed by
four independent observations (measurements of: persistence time vs
glycerol/water concentration, persistence time vs glycerol/ethanol
concentration, extinction rates at different times vs
glycerol/water concentration and persistence time vs temperature).

The extinction rates of the pipes are always exponentially
decreasing with time elapsed since formation, but different time
scales for the evolution of the phenomenon have been observed, the
appearance of which depends on the concentration of the liquid
used.

\section{Theoretical insight}

\noi Our experimental analysis has shown a very rich phenomenology
for the pipe effect, that cannot be simply accounted for standard
relaxation effects. Although spurious effects, such as the
dissolution of small air bubbles in the liquid, are sometimes
present, we have seen that the primary phenomenon of the formation
of a well-defined structure is not at all related to those
secondary effects. Also, the possible dissolution of microscopic
air bubbles located in the interstices of the surface of the metal
sphere (these bubbles cannot be resolved with our experimental
apparatus), as the explanation for the alteration of the
properties of the liquid encountered by the falling body, is
largely excluded by the size of the pipe formed and by the
non-trivial phenomenology observed. Then, we have to search for a
more complex theoretical explanation.

\begin{table}
\begin{center}
\footnotesize
\begin{tabular}{|r|l|l|l|}
\hline
 & $t_c$ (min) & $T_0$ ($ ^{\rm o}$C) & $T_\sigma^2$ ($ ^{\rm
 o}$C$^2$)
\\ \hline
1. & $11.0 \pm 0.2$ & $36.0 \pm 0.3$ & $49 \pm 8$  \\
\hline
2. & $157 \pm 2$ & $19.6 \pm 0.3$ & $90 \pm 5$  \\
\hline
\end{tabular}
\caption{Fitting parameters for the persistence time curves in Eq.\,(\ref{eqperst}) (see the text). \newline} \label{tableperstemp}
\end{center}
\vspace{-0.7truecm}
\end{table}

As a first step in the understanding of the phenomena reported
above, let us assume that the falling of an heavy sphere in
glycerol or other viscous media induces, substantially, only the
formation of a {\it separation surface} between the bulk liquid
and that in direct contact with the falling body. Possible slight
alteration of the liquid properties, as effectively observed, will
be not primarily taken into account for the moment.

Two main processes should, then, receive at least a rough
explanation: the formation of such a surface and the time
evolution of the pipe. These two items will
be considered in the next paragraphs in the most simple way,
leaving details (even important) for possible future studies.

\subsection{A microscopic model for pipe formation}

\noi The development of a surface in polar liquids, such as water,
glycerol, etc., is characterized by a well definite orientation of
the molecular dipole moments on the said surface
\cite{physicalchemistry} \cite{bottcher} \cite{croxton}: the
degree of polarization of the surface molecules is, effectively,
greater than that of bulk ones. On the surface of the liquid, a
net polarization of the molecular dipole moments develops, with an
associated generation of a non-vanishing surface electric field
and potential. Such polarization is, evidently, due to the
presence of a polarizing electric field that for water, for
example, derives from the non-vanishing permanent quadrupole
moment of the water molecule. The polarizing electric field
generates a torque on the surface molecules that tends to
orientate the dipole moments along a given direction but, of
course, a transition zone (that can extends even over $10 \div
100$ molecular layers) exists before reaching the true bulk. More
specifically, in liquids characterized by the presence of hydrogen
bonds, domains can form where the dipoles are oriented along the
direction of maximum polarization for the given domain, while the
direction of maximum polarization of neighboring domains tend to
rotate of $180^{\rm o}$ with respect to the previous one
(similarly to the juxtaposition of north and south poles for
magnets). The presence of an electric field, then, increases the
number of ``favorably'' oriented domains at the expenses of the
those less ``favorably'' ones, until a stationary state is
reached.

Here we assume that the development of the pipe surface occurs in
a way similar to the standard one described above. Very simply,
the polarizing electric field may be thought as generated by the
``friction'' between the falling sphere and the viscid liquid,
whose value is completely non-negligible due to the high viscosity
of the liquids studied. An estimate of the order of magnitude for
the field generated in such a way may be roughly obtained as
follows.

The energy available from the falling sphere is that of the
gravitational field, accounting for\footnote{Here and in the
following, for our numerical estimates we use typical values as
used in our experimental setup; for example we take spheres of
radius $R \approx 1$ mm falling in glycerol from an height of
$\approx 10$ cm, etc.}:
\be \label{eqgrav} %
U_G = m_{\rm sphere} \, g \, h \simeq 3 \times 10^{-5} {\rm J}.
\ee %
A part of such energy will be dissipated in the production of heat
(that is, it will increase just the thermal agitation of the
liquid) due to the slowing down action of the liquid on the
falling sphere. This energy may be calculated as the work $U_{\rm S}$
done by the Stokes force $F=6 \pi R \eta v$ during the falling:
\be \label{eqslowing}%
U_{S} \simeq 6 \pi R \, \eta \, v_L^2 \, \vartheta \simeq 5
\times 10^{-6} {\rm J}
\ee %
[where $v_L = 2(\rho_{\rm sphere}-\rho_{\rm liquid})g R^2/9 \eta$
is the terminal velocity and $\vartheta = 2 \rho_{\rm sphere}
R^2/9 \eta$ is the time constant of the instantaneous velocity of
the sphere].

\begin{figure}
\begin{center}
\begin{tabular}{ccc}
\epsfxsize=3.5cm %
\epsfysize=3cm %
\epsffile{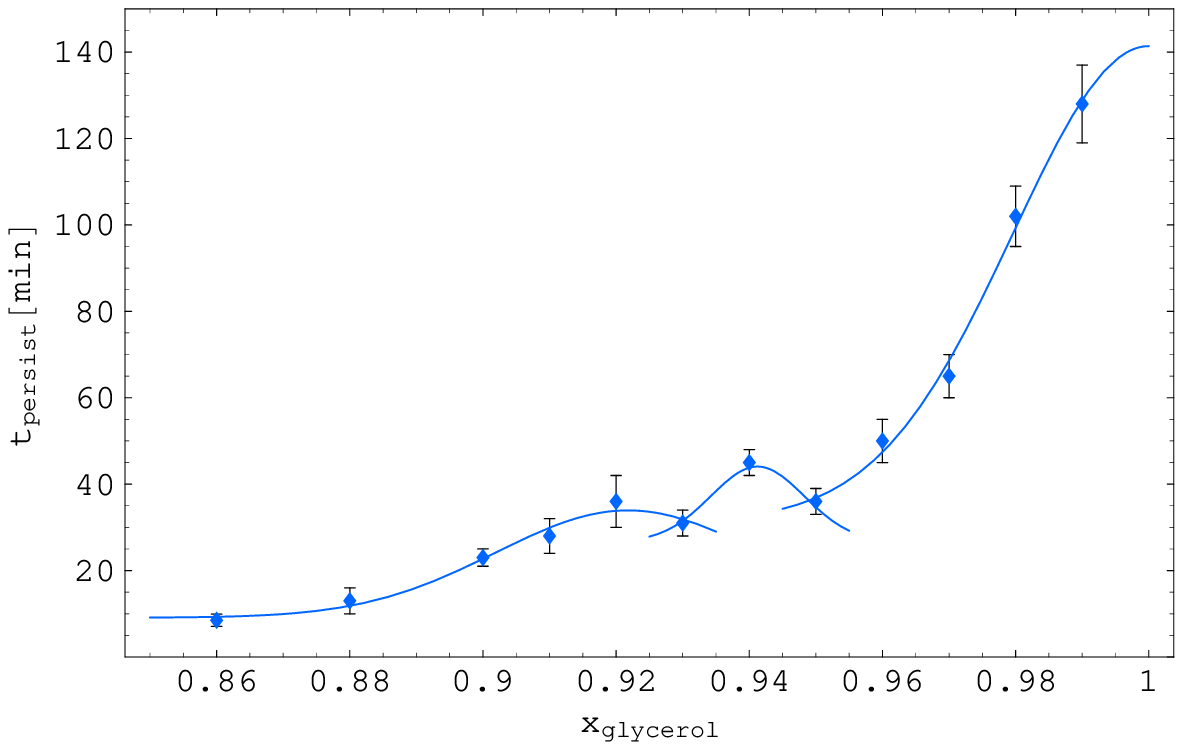} a) & &
\epsfxsize=3.5cm %
\epsfysize=3cm %
\epsffile{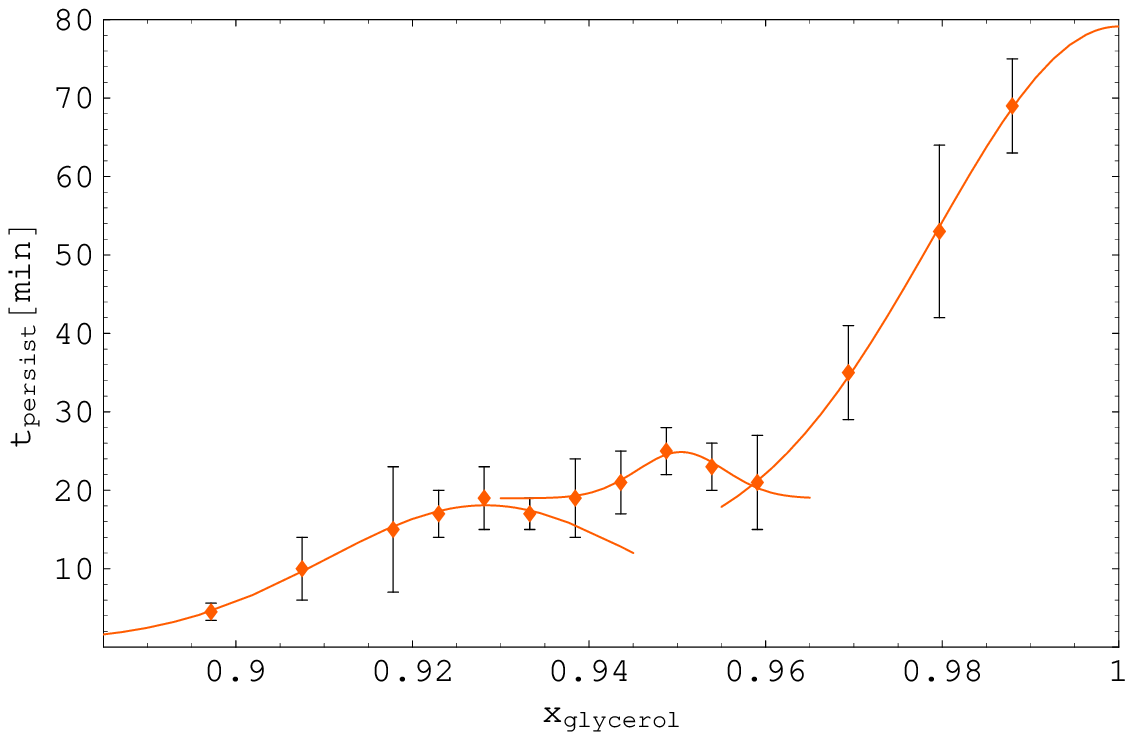} b)
\end{tabular}
\caption{Measured persistence times for pipes generated in
glycerol water (a) and glycerol/ethanol(b) mixtures as functions of
the glycerol concentration have been fitted by three different
gaussian curved (see the text). \newline} \label{figfitpersist}
\end{center}
\vspace{-0.7truecm}
\end{figure}

We assume that a non-vanishing energy $U_P$ available for the
polarization mechanism described above exists, whose order of
magnitude is given by:
\be \label{eqpol} %
U_P \simeq U_G - U_{S} \approx 10 ^{-5} {\rm J}.
\ee %
Such energy will be used to orientate the molecular electric
dipoles, so that it will be proportional to the polarizing field
$E$:
\be \label{eqdipole} %
U_D = n \, \mu \, E,
\ee %
where $\mu$ is the electric dipole moment of the glycerol molecule
and $n$ the number of electric dipoles, which is {\it
approximatively} given by the number of molecules intercepted by
the falling sphere. The number of molecules in 1 g of glycerol is
$n_0 = N_A/M \simeq 6.5 \times 10^{21}/ {\rm g}$ ($N_A$ is the
Avogadro number, while $M$ the molecular weight of glycerol). For
simplicity we assume that the pipe formed is a cylinder, so that
$m_{\rm pipe} = \rho_{\rm liquid} \pi R^2 h \simeq 0.4 {\rm g}$
and $n = n_0 m_{\rm pipe} \simeq 2.5 \times 10^{21}$ molecules. By
equating Eq.\,(\ref{eqpol}) to Eq.\,(\ref{eqdipole}) and using the
values in Table \ref{tableprop}, we finally get:
\be \label{eqfield} %
E \approx 450 \, {\rm V/m},
\ee %
which is a fairly optimistic estimate, since $U_G - U_{S}$ may be
even lower by one order of magnitude (if the falling time is much
greater than the time constant $\vartheta$). Anyway, such an
estimate completely agrees with the order of magnitude of the
classic Frenkel estimate \cite{frenkel} of the surface potential
of water, so that it is quite useful, in the following, to explore
further the approximate microscopic model envisaged above.

In such scenario, it is very likely that the liquid molecules in
the pipe and in the bulk near it experience a non-homogeneous
electric field whose maximum amplitude is reached just on the pipe
surface. It expectedly decreases away from the surface inside the
pipe, while the field is practically zero in the bulk, where the
``friction'' action by the sphere is absent.

A non-homogeneous electric field \mbox{\boldmath $ E$} exerts a translational
force \mbox{\boldmath $F$} on a given molecule,
\be \label{eqfe} %
\mbox{\boldmath $F$} = \left( \mbox{\boldmath $\mu$} \!\cdot\! \mbox{\boldmath
$\nabla$} \right) \mbox{\boldmath $E$} + \alpha \left(\mbox{\boldmath $E$} \!\cdot\!
\mbox{\boldmath $\nabla$} \right) \mbox{\boldmath $E$}
\ee %
(\mbox{\boldmath $\mu$} and $\alpha$ being the electric dipole
moment vector and the polarizability of the liquid molecule,
respectively), so that the molecules with a dipole moment pointing
along the direction of \mbox{\boldmath $ E$} will be pushed towards the regions
with a greater field intensity. In such a way, a concentration
gradient generates which induces a decrease of the density inside
the pipe, with an effect similar to that of electrostriction. From
the standard theory of such effect \cite{bottcher} we can evaluate
the variation $\Delta \rho = \rho_{\rm bulk} - \rho_{\rm pipe}$ of
the density between the pipe and the bulk, obtaining:
\be \label{eqdeltarho} %
\frac{\Delta \rho}{\rho_{\rm bulk}} \simeq \frac{E^2}{8 \pi} \,
\beta \rho_{\rm bulk} \left( \frac{\partial \varepsilon}{\partial
\rho} \right)_T \equiv \left( \frac{E}{E_0} \right)^2,
\ee %
where $\beta$ and $\varepsilon$ are the isothermal compressibility
and dielectric constant of the (bulk) liquid. As a simplifying
assumption, we take valid \cite{bottcher} the Debye theory of
electric polarization for calculating the variation of the
dielectric constant with the density,
\be \label{eqdieldens} %
\left( \frac{\partial \varepsilon}{\partial \rho} \right)_T \simeq
\frac{3}{M} \, \frac{\varepsilon -1}{\varepsilon + 2}.
\ee %
From the values in Table \ref{tableprop} we get the following
estimate for the typical field $E_0$ defined in Eq.\,(\ref{eqdeltarho}):
\be \label{eqe0} %
E_0 \approx 260 \, {\rm V/m},
\ee %
which is of the same order of magnitude of the field in Eq.\,(\ref{eqfield}).

By assuming, as suggested by the preliminary experimental data
(\ref{visc}), that
\[
\frac{\Delta \rho}{\rho_{\rm bulk}} \approx \frac{\Delta
\eta}{\eta_{\rm bulk}} \simeq 4 \times 10^{-2} ,
\]
we deduce that the electric field required to generate such a
variation of the density, according to the mechanism described,
would be:
\be \label{eqee0} %
E \approx 0.2 E_0 \approx 50 \, {\rm V/m},
\ee %
an order of magnitude estimate in evident agreement with the
independent one obtained above in Eq.\,(\ref{eqfield}).

A slight increase of the liquid density near the pipe surface,
with an associated decrease inside the pipe, due to presence of
the polarizing electric field, implies a corresponding change
$\Delta \varepsilon$ of the dielectric constant of the liquid in the
pipe with respect to the bulk, as anticipated by the formulae
above. From the standard theory \cite{bottcher} we have:
\be \label{eqdeltadiel} %
\Delta\varepsilon \simeq \frac{E^2}{4 \pi} \, \beta \, \rho_{\rm
bulk}^2 \left( \frac{\partial\varepsilon}{\partial \rho} \right)_T^2
\simeq \frac{6}{M} \, \frac{\varepsilon -1}{\varepsilon + 2} \, \left(
\frac{E}{E_0} \right)^2
\ee %
(within the mentioned approximations). For $(E/E_0)^2 = \Delta
\rho / \rho_{\rm bulk} \simeq 4 \times 10^{-2}$ we get:
\be \label{eqdeldiel} %
\Delta\varepsilon \approx 2 \times 10^{-3}.
\ee %
The variation of the dielectric constant between the pipe and the
bulk is, thus, negligible, so that the optical properties of the
pipe are practically equal to those of the bulk. This, however,
does not apply to the interface, i.e. on the pipe surface, where a
gathering of electric dipoles is present. In all respects, the
pipe behaves as a (almost cylindrical) dielectric shell, as indeed
observed experimentally.

\subsection{A thermodynamical model for pipe evolution}

\noi Once the surface of the pipe has been formed by the falling
of a sphere, the subsequent evolution consists just in the
thinning of the pipe itself. More precisely, according to our
observations (see the points E2-E3), we may distinguish the main,
most relevant, process of pipe narrowing followed by a final
``dissolution'' of the pipe, which gets deformed. In the narrowing
process the pipe keeps its structure, while it ``dissolves'' once
the pipe radius reduces to small fractions of its initial value.
Here we focus our attention on the main narrowing process, which
we will show it can be described by standard thermodynamical
procedures.

\begin{figure}
\begin{center}
\epsfxsize=8cm %
\epsffile{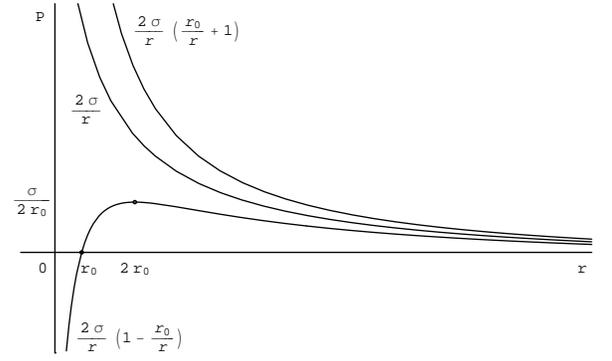} %
\caption{The different solutions of Eq.\,(\ref{eqtherm2})
according to the values of the integration constant $y_0$.
\newline} \label{plotp}
\end{center}
\vspace{-0.7truecm}
\end{figure}

As a starting point, we assume that the narrowing process takes
place at approximately constant pressure (that of the bulk liquid)
and that no appreciable heat transfer occurs with the
surroundings. In other words, the bulk liquid is thought as a
pressure reservoir, inside which the narrowing of the pipe occurs
adiabatically. In such approximations, the work made to narrow the
pipe is stored as enthalpy of the system: the change of enthalpy
$\drm H$ for an almost isobaric, adiabatic change of state is the
utilizable work $\drm W$ gained by the system aside from the
volume work \cite{thermo}. This is essentially the work done by
the liquid to induce a variation $\drm A$ of the pipe surface,
which is equal to $\sigma \drm A$ ($\sigma$ being the surface
tension of the viscous liquid, whose value is supposed to be equal
in the bulk and in the pipe). As a simplifying assumption, we will
also consider that the internal energy of the pipe does not change
while it evolves, so that the change of enthalpy is given by $P
\drm V + V \drm P$, $V$ being the pipe volume and $P\equiv P_{\rm bulk}
- P_{\rm pipe}$ the net pressure. The condition above reads, then:
\be \label{eqtherm1} %
P \drm V + V \drm P \simeq \sigma \drm A .
\ee %
The interpretation of such equation is straightforward: the energy
spent in reducing the pipe surface ($\sigma \drm A$) equals the
work done in reducing the pipe volume ($P \drm V$) plus the energy
to contrast the variation of the bulk reservoir ($V \drm P$),
similarly to what happens in the classical Joule-Thomson effect
(where, however, no surface variation occurs, so that $\drm H =
0$). For simplicity we consider the pipe as a cylinder of radius
$r$ and height $h$ much greater than the radius, so that $\drm V =
2 \pi h \, r \, \drm r$ and $\drm A \simeq 2 \pi h \, \drm r$. By
substituting in Eq.\,(\ref{eqtherm1}), we have:
\be \label{eqtherm2} %
r^2 \frac{\drm P}{\drm r} + 2 \left( r P - \sigma \right) = 0.
\ee %
The solution of this differential equation\footnote{It may be
easily obtained by the change of variable $P \rightarrow y \equiv
r^2 P / \sigma$ ($y$ has the dimensions of a length).} is the
following:
\be \label{eqtherm3} %
P = \frac{2 \sigma}{r} \left( 1 \pm \frac{r_0}{r} \right),
\ee %
where $r_0$ is a (positive) integration constant with the
dimensions of a length. As shown in Fig.\,\ref{plotp}, the solution
$P(r)$ depends crucially on the sign above; if the negative sign
is taken, $P(r)$ has a relative maximum and goes to zero for $r =
r_0$.

From the physical viewpoint, some considerations are in order. If
the positive sign is considered in Eq.\,(\ref{eqtherm3}), the
difference $P_{\rm bulk} - P_{\rm pipe}$ is positive for any
radius of the pipe, and indefinitely increases for $r$ approaching
zero. Then, in such a case,
the pipe expands, this situation not corresponding to what
observed. Instead, If the negative sign is taken in
Eq.\,(\ref{eqtherm3}), the difference $P_{\rm bulk} - P_{\rm
pipe}$ is positive for large radii and increases till a maximum;
after that, the net pressure decreases, reaching the equilibrium
condition $P_{\rm bulk} = P_{\rm pipe}$ for $r = r_0$, and becomes
indefinitely negative for $r$ approaching zero. Actually, the
narrowing of the pipe corresponds to the last case, when its
radius decreases from the maximum value $2r_0$ to the value $r_0$.

We can then envisage the following picture. At the very beginning,
just after the falling of the sphere, when the pipe undergoes a
rapid thinning (see the point D1), the surface ``compresses'' the
liquid inside the pipe, inducing an increase of the internal
pressure ($P_{\rm pipe}$) till a maximum $P_{\rm max} = \sigma / 2
r_0$ for $r=2 r_0$. By introducing the numerical values of Tables
\ref{tableprop} and \ref{tabledratio}, we have $P_{\rm max}\approx
300$ Pa, which is a reasonable small value (in the present model)
for the pressure difference. This first region corresponds to the
balance of the surface effects with the volume ones, $\sigma \drm
A = P \drm V$ (that is, $V \drm P \simeq 0$), for which $P \simeq
\sigma /r$. By decreasing the radius $r$, the volume effects
become subdominant with respect to the surface ones, while the
role of the energy contrasting the variation of the bulk reservoir
is no more negligible. This continues until $P_{\rm bulk} = P_{\rm
pipe}$ for $r=r_0$, and such equilibrium point may be considered
as ``conclusive'' of the narrowing process, since for $r < r_0$ we
have $P_{\rm pipe} > P_{\rm bulk}$.

The present model, then, predicts naturally the appearance of a
length scale $r_0$, which can be directly related to the observed
value $d_0$ in Eq.\,(\ref{eqdratio}). Indeed, on one hand, $r =
r_0$ is the value of the pipe radius at which the narrowing
process ``ends''; on the other hand, $d_0 \cdot r_{\rm sphere}$
(with $d_0$ in Eq.\,(\ref{eqdratio})) is just the value of the
pipe radius for $t \rightarrow +\infty$, so that we simply have:
\be \label{r0d0} %
r_0 = d_0 \cdot r_{\rm sphere} . 
\ee %
The phenomenological interpretation of the parameter $t_0$, that
is the ``initial'' time \footnote{Note that, according to
observations, the present model would apply only to the
``stationary'' pipe evolution, while other effects occurring just
at the formation of the pipe or soon after that are not included.
Thus, it is practically meaningless to consider the part of the
$P(r)$ curve for $r > 2 r_0$.} of the narrowing process, as
deduced from experimental observations, is as well straightforward
in the present model. In fact, from Eq.\,(\ref{eqdratio}) it
follows immediately that the time $t_0$ corresponds to the point
$d_{\rm ratio} = 2 d_0$ and, from what said above, for $r = 2 r_0$
the net pressure $P_{\rm bulk} - P_{\rm pipe}$ takes its maximum.
We then have:
\be \label{t0pmax} %
\left. P \right|_{t=t_0} = P_{\rm max} = \frac{\sigma}{2 r_0}
\ee %
The appearance %
of different phases in the time evolution of the pipe, as observed
in Fig.\,\ref{figdratio} (see the point R2), may be described in
the present scheme in terms of a sudden change of the net
pressure. Indeed, $r_0$ decreases from one phase to the following
(see the values of $d_0$ in Table \ref{tabledratio}), so that from
Eq.\,(\ref{t0pmax}) an increase of $P_{\rm max}$ follows. The
change of the time scale during the pipe evolution is, then,
caused by an increase of the net pressure or, more specifically,
by a decrease of the pipe pressure.

The evolution of $P$ with time $t$ may be obtained by inserting
Eq.\,(\ref{eqdratio}) into Eq.\,(\ref{eqtherm3}) (with the minus
sign):
\be \label{eqtherm4} %
P = \frac{2 \sigma}{r_0} \, \frac{\displaystyle \erm^{\frac{t -
t_0}{\tau}}}{\displaystyle \left( 1 + \erm^{\frac{t - t_0}{\tau}}
\right)^2} .
\ee %
The curve $P(t)$ is, thus, symmetric with respect
to $t=t_0$. Near its maximum, that is at the start of pipe
narrowing, it is approximated by a gaussian centered around $t_0$
and with full width at half maximum equal to $2 \tau$:
\be \label{eqtherm5} %
P \simeq \frac{\sigma}{2 r_0} \, \erm^{- \frac{(t- t_0)^2}{4
\tau^2}}.
\ee %
Instead, at the end of the narrowing (for $r$ approaching $r_0$),
the net pressure decreases exponentially with a time constant
$\tau$:
\be \label{eqtherm6} %
P \simeq \frac{2 \sigma}{r_0} \, \erm^{- \frac{t- t_0}{\tau}}.
\ee %
An increase of $P$ from one phase to the following then
corresponds to an increase of $t_0$ and/or $\tau$, in agreement
with the experimental data (see Table \ref{tabledratio}).

Summing up, we find that pipe evolution is perceptible when
$P_{\rm max}$ is sensibly different from zero, that is for quite
large values of $\sigma$ and/or $r_0$ as, indeed, effectively
observed. In general, the effect studied is ruled {\it only} by
these two parameters. We also point out that, in glycerol/water
mixtures, an increase of the water content leads to an increase of
the surface tension $\sigma$, so that we have an increase of
$P_{\rm max}$. Since, according to Eq.\,(\ref{eqtherm4}), the
difference between bulk and pipe pressure increases, the narrowing
process is thus faster for water-rich mixtures, as it is generally
apparent in Figs.\,\ref{figdratio} or \ref{figfitpersist}. The
interpretation of the maxima in Figs.\,\ref{figfitpersist}, as well
as other peculiar observations, is however beyond the range of the
present thermodynamical model, and claims for a more detailed
microscopic interpretation of the phenomenon.

\section{Conclusions}

\noindent In this paper we have studied in detail the pipe effect
occurring in viscous liquids which, despite its simplicity, at the
best of our knowledge it has not been reported previously in the
literature. The effect, observed here in pure glycerol, in
glycerol/water and glycerol/ethanol mixtures, and in castor oil,
consists in the formation of a well-defined structure (a pipe)
after the passage of a heavy body in the viscous liquid, this
structure lasting for quite a long period.

A very rich phenomenology has been observed for such effect, that
cannot be trivially explained in terms of standard relaxation
and/or spurious effects like the dissolution of air bubbles in the
part of liquid encountered by the falling body.

A part of this phenomenology may be well summarized by simply
admitting that the pipe effect is primarily due to the formation
of a separation surface, upon which mechanical action of even
different kind may be performed. The apparently non-obvious
optical effects observed are, then, just the manifestation of this
surface to light (normal or laser) impinging on it.

The time evolution of the present phenomenon is, as well, not at
all trivial, as our quantitative measurements of the extinction
rates and persistence time curves of the pipe have shown. While
the pipe gets thinner and thinner as the time goes on, it
experiences different exponentially decreasing time phases, with
different time scales, this effect being likely due to sudden
changes of the liquid pressure inside the pipe. By lowering the
viscosity of the sample liquid, the persistence time of the pipe
gets shorter, till quite a complete non-observability of the
effect for not high values for the viscosity. This transition has
been accurately studied by measuring the persistence time curves
for glycerol/water (or glycerol/ethanol) mixtures with decreasing
content of glycerol. As expected, the persistence time generally
decreases with decreasing concentration of glycerol but, quite
interestingly, two maxima have been observed for given
concentrations, this behavior being confirmed by several
independent observations.

Our study has not been limited to experimental observations, but
we have also tried to construct a theoretical model for the
formation and the evolution of the pipe, able to interprete (at
least a part of) the data found.

We have, thus, assumed that the falling body induces a
non-negligible polarizing electric field in highly viscous
liquids. It tends to orientate the molecular dipole moments of the
liquid (the samples employed are highly polar liquids, with the
presence of hydrogen bonds) in the neighborhood of the falling
body surface, thus generating a dielectric shell (the separating
liquid surface) that would be responsible of (part of) the
phenomenology observed. Indeed, order of magnitude estimates of
the polarizing electric field, for the case studied here, show
that such picture could not be far from reality, since the model
brings to definite predictions that have been effectively
observed.

The dynamical evolution of the pipe, once formed according to the
microscopic model proposed, seems, then, well described by a
thorough thermodynamical model. According to this, the evolution
is just the macroscopic result of the balance of two competing
actions: the expansion of the pipe surface ruled by the surface
tension effect, contrasted by the bulk liquid that is assumed to
be an almost perfect pressure reservoir. This model gives a likely
account of the experimentally observed appearance of a length
scale $r_0$ in the extinction rate of the pipe. Together with the
surface tension parameter, it rules the behavior of the net
pressure acting on the pipe surface, which reaches a maximum at
the onset of the narrowing process, the ``initial'' time $t_0$
being directly obtained from measurements. Several other
predictions come out from the model studied, fitting quite well
with the experimental observations.

Although some understanding of the effect observed
has been apparently reached, it should be
regarded mainly as preliminary, since further experimental
investigations are needed. Also, the interpretative models
proposed for the formation and evolution of the pipe should be
considered, as well, just as a starting theoretical framework,
which claims for a more detailed inspection. Novel results,
from both theory and experiments, are then likely to be expected.

\vspace{1truecm}

\begin{acknowledgments}
\noi The present study was inspired by some observations from a
student of one of us (S.E.), Riccardo Frasca; our gratitude to him
and to Chiara Baldassarri for essential experimental assistance is
here acknowledged. Moreover, it would not have been possible to
carry out accurately the experiments discussed in this paper
without the valuable help of Ercole Gatti, Marco Marengo and his
group (Alfio Bisichini, Stefano dall'Olio, Carlo Antonini),
Massimo Lorenzi, Giovanna Barigozzi and Giuseppe Rosace.
\end{acknowledgments}



\begin{thebibliography}{}

\bibitem{HBond1}
P.K. Dixon, L. Wu and S.R. Nagel, Phys. Rev. Lett. {\bf 65} (1990)
1108; W. G\"otze and L. Sj\"ogren, Rep. Prog. Phys. {\bf 55}
(1992) 241; P. Lunkenheimer, A. Pimenov, M. Dressel, Yu.G.
Goncharov, R. B\"ohmer and A. Loidl, Phys. Rev. Lett. {\bf 77}
(1996) 318; C. Hansen, F. Stickel, T. Berger, R. Richert and E.W.
Fischer, J. Chem. Phys. {\bf 107} (1997) 1086; C.A. Angell, K.L.
Ngai, G.B. McKenna, P.F. McMillan and S.W. Martin, J. Appl. Phys.
{\bf 88} (2000) 3113;  S. Sudo, M. Shimomura, N. Shinyashiki and
S. Yagihara, J. Non-Cryst. Solids {\bf 307} - {\bf 310} (2002)
356; Th. Blochowicz, Ch. Tschirwitz, St. Benkhof and E.A.
R\'ossler, J. Chem. Phys. {\bf 118} (2003) 7544; A. Sanz, M.
Jimenez-Ruiz, A. Nogales, D. Martýn y Marero and T.A. Ezquerra,
Phys. Rev. Lett. {\bf 93} (2004) 015503; Li-Min Wang, S. Shahriari
and R. Richert, J. Phys. Chem. {\bf B 109} (2005) 23255.

\bibitem{HBond2}
E. Donth, The Glass Transition: Relaxation Dynamics in Liquids and
Disordered Materials (Springer, Berlin, 2001); F. Kremer and A.
Schonhals (Eds.), Broadband Dielectric Spectroscopy; (Springer,
Berlin, 2003);

\bibitem{Glyc12}
J.W. Lawrie, Glycerol and the glycols: Production, properties and
analyses (Chemical Catalog Company, New York, 1928); C.S. Miner
and N.N. Dalton, Glycerol (Reinhold, Baltimore, 1953); A.R.
Ubbelohde, The Molten state of matter (Wiley, New York, 1978;
Chap. 16).

\bibitem{cryopro}
R.E. Lee, Jr., C.P. Chen, and D.L. Denlinger, Science {\bf 238}
(1987) 1415; F. Franks (Ed.), Water: A comprehensive treatise
(Plenum, New York, 1982, Vol. 7).

\bibitem{cryopre}
P. Mazur, Science {\bf 168} (1970) 939; G.M. Fahy, D.I. Levy and
S.E. Ali, Cryobiology {\bf 24} (1987) 196; Z. Chang and J.G.
Baust, {\it ibid.} {\bf 28} (1991) 268; G. Vigier and R.
Vassoille, {\it ibid.} {\bf 24} (1987) 345; P. Boutron and F.
Arnaud, {\it ibid.} {\bf 21} (1984) 348.

\bibitem{glassformer}
C.A. Angell, in K.L Ngai and G.B. Wright (Eds.), Relaxations in
complex systems (NRL, Washington D.C., 1985); R. B\'ohmer and C.A.
Angell, Phys. Rev. {\bf B 45} (1992) 10091.

\bibitem{glycerolprop}
See the websites: \\
{\footnotesize
http://www.dow.com/glycerine/resources/physicalprop.htm}, \\
{\footnotesize
http://www.thegoodscentscompany.com/data/rw1008461.html}, \\
{\footnotesize http://en.wikipedia.org/wiki/Glycerol}.

\bibitem{waterprop}
See the website {\footnotesize
http://www.lsbu.ac.uk/water/data.html} and references therein.

\bibitem{physicalchemistry}
H. Eyring, D. Henderson and W. Jost (eds.), Physical chemistry: An
advanced treatise, Vol. VIIIA - Liquid state, edited by D.
Henderson (Academic Press. New York, 1971).

\bibitem{bottcher}
C.J.F. Bottcher, Theory of electric polarization (Elsevier,
Amsterdam, 1973)

\bibitem{croxton}
C.A. Croxton, Statistical mechanics of the liquid surface (Wiley,
New York, 1980).

\bibitem{frenkel}
J. Frenkel, Kinetic theory of liquids (Dover, New York, 1955).

\bibitem{thermo}
L.E. Reichl, A modern course in statistical physics (Wiley, New
York, 1998, 2nd edition), pp.40 ff.; W. Greiner, L. Neise and H.
St\"ocker, Thermodynamics and statistical mechanics (Springer, New
York, 1995), pp.95 ff.


\end{thebibliography}
\end{document}